\documentclass{aa}

\usepackage{graphicx}
\usepackage{txfonts}
\usepackage{bm}
\usepackage{cases}

\usepackage{hyperref}   
\hypersetup{colorlinks=true,linkcolor=blue,citecolor=blue,filecolor=blue,urlcolor=blue}

\begin{document}

\title{Ring formation and dust dynamics in {wind-driven} protoplanetary discs: {global simulations}}
\titlerunning{Ring formation and dust dynamics in PP discs}

\author{A. Riols\inst{1},  G. Lesur\inst{1} \and F. Menard\inst{1}}
\institute{%
$^1$ Univ. Grenoble Alpes, CNRS, Institut de Planétologie et d’Astrophysique de Grenoble (IPAG), F-38000, Grenoble, France
}

\date{\today}

\abstract{{Large-scale vertical magnetic fields are believed to play a key role in the evolution of protoplanetary discs. Associated with non-ideal effects, such as ambipolar diffusion, they are known to launch a wind that could drive accretion in the outer part of the disc ($R> 1$ AU). They also potentially lead to self-organisation of the disc into large-scale axisymmetric structures, {similar to} the rings recently imaged by sub-millimetre or near-infrared instruments (ALMA and SPHERE). } 
The aim of this paper is to investigate the mechanism behind the formation of these gaseous rings,  but also to understand the dust dynamics and its emission in discs threaded by a large-scale magnetic field. {To this end}, we {performed} global magneto-hydrodynamics (MHD) axisymmetric simulations with ambipolar diffusion using a modified version of the PLUTO code.  We {explored} different magnetisations with the midplane $\beta$ parameter ranging from $10^5$ to $10^3$ and included dust grains -treated in the fluid approximation- ranging from $100 \mu$m to 1 cm in size. {We} first {show} that the gaseous rings (associated with zonal flows) {are tightly linked to the existence of MHD winds}.  { Secondly, we find that} {millimetre-size dust} is highly sedimented, with a typical scale height  of 1 AU at $R=100$ AU for $\beta=10^4$,  compatible with recent ALMA observations.  We also show that {these grains} concentrate  into pressure maxima associated with zonal flows, leading to the formation of dusty rings. Using the radiative transfer code MCFOST, we computed the dust emission and make predictions on the ring-gap contrast and the spectral index that one might observe with interferometers like ALMA.}


\maketitle


\section{Introduction}
Protoplanetary discs, the nurseries of planets,  are known to dissipate over a few million years. Measurements of UV excess emission indicate that these  discs are accreting with rates 
ranging from $10^{-10}$ to $10^{-7}$ solar mass per year onto the central object \citep{venuti14}. The accretion process plays a central role in the formation of stars, but also in discs and planets evolution.  However, its origin is still actively debated. For several decades, accretion was believed to result from turbulent motions, which act like an effective viscosity in the discs \citep{shakura73} transporting the angular momentum outwards.  The most commonly invoked excitation mechanism of turbulence in accretion discs is the magneto-rotational instability  \citep[MRI, ][]{balbus91,hawley95}. Several MHD simulations in the ideal limit showed that the MRI could provide the desired accretion rate and transport efficiency \citep{flock11,flock13,bai13}. However, 
due to their optical thickness and low temperatures, protoplanetary discs are weakly ionised in their midplane beyond 0.1–1 au  \citep{gammie96}. As a result,  non-ideal MHD effects, such as ohmic diffusion, Hall effect, and ambipolar diffusion dominate at these radii and tend to suppress any form of MRI-driven turbulence \citep{fleming00, sano02,wardle12,bai13b,lesur14, bai15}.

Recent observations at sub-millimetric wavelength with ALMA seem to confirm that the level of turbulence in protoplanetary discs is indeed very low.  Firstly, there is some evidence that the settling of millimetre dust is quite strong in some discs, like HLTau  \citep{pinte16}, which excludes the presence of vigorous turbulent motions in these discs. Secondly, the measurements of gas velocity dispersion in CO  \citep[e.g.][]{flaherty15,Flaherty17}  indicate that the departure from the Keplerian velocity in a number of discs is much weaker than what was theoretically predicted from ideal MRI simulations. Yet, these discs are known to accrete, which suggests that the transport of angular momentum is essentially driven by laminar processes. \\

Quite apart from the accretion problem, observations have revealed the presence of many sub-structures in several systems, both in the dust continuum (e.g. with ALMA), and in the scattered light emission (e.g with SPHERE). In particular, one of the most remarkable features is the concentric rings (or gaps) that appear in 60 to 70 \% of the observed discs  \citep{garufi18}. Some examples of discs showing rings are HL tau \citep{alma15},  TW Hydra \citep{andrews16}, or the disc around Herbig Ae star HD 163296 \citep{isella16}. These structures are possibly altering the long-term disc evolution and could be privileged locations of dust accumulation \citep{pinilla12}, a key step towards planetary core formation. How these rings form remains an open question, though a number of physical mechanisms have been proposed so far.  The most commonly invoked scenario is the presence of planets opening gaps in the disc \citep{dong15, dipierro15}. Other scenarios involve  snow lines \citep{okuzumi16}, dust-drift-driven viscous ring instability \citep{wunsch05,dullemond18},  MHD turbulence with dead zones \citep{flock15}, zonal flows \citep{johansen09},  or  secular gravitational instabilities in the dust \citep{takahashi14}. 

In summary, the fundamental question is: are these rings the result of planet formation (as the planet-induced scenario suggests) or the driver (as envisioned by purely gas-induced mechanisms)? We note that answering this question is more difficult than it seems, since in both cases, one eventually expects to see planets in the core of the gaps. Hence, finding planets in these gaps cannot be the key signature allowing one to distinguish between these scenarios. Instead, one should rely on specific signatures of planet-free rings and gaps, which can be tested on the sky. \\

It has been recently suggested that magnetic winds in discs could play a crucial role in both the accretion and the ring formation processes.  Molecular outflows consistent with magnetic winds have been observed  in several Class 0 and I discs \citep{Launhardt09,lumbreras14, Bjerkeli16,tabone17,hirota17,louvet18}, although their origin is still debated.   Simulations including a large-scale poloidal field and non-ideal effects (Hall effect, ambipolar diffusion, mainly) have shown that discs are able to launch a wind from their surface removing angular momentum and driving accretion in the process \citep{bai13b,lesur14,simon15,gressel15,bethune17}. The accretion is made possible by the laminar torque associated with the large-scale magnetic field supporting the wind. The typical accretion rates measured in these simulations are compatible with the observations. Moreover,  large-scale axisymmetric structures  or `zonal flows' associated with rings of matter \citep{kunz13,bai15,bethune16,bethune17,suriano18} form spontaneously in these non-ideal simulations. The ring formation has been attributed to several different MHD processes: self-organisation by ambipolar diffusion \citep{bethune17}, reconnection of poloidal fields lines \citep{suriano18,suriano18b}, or  anti-diffusive processes \citep{bai14}. More recently however,  \citet{riols19} argued that their formation could result from a more generic process involving a wind instability and operating in both ideal and non-ideal MHD. 
 
One challenging problem is to make a bridge between these non-ideal simulations and observations. Since current sub-millimetre or near-infrared observations by ALMA and SPHERE have mostly traced the dust emission so far (since the gas is more difficult to detect), it is crucial to understand the dynamics of dust grains in non-ideal MHD flows. In particular, there is concern surrounding the impact of gaseous rings  and a large-scale wind on their spatial distribution. Several  simulations by \citet{zhu15,xu17} have combined the dust dynamics with ambipolar diffusion but focused on a large particle regime  ($\gtrsim$ mm). More recently, \citet{riols18} studied the settling and dynamics of smaller grains in ambipolar-dominated flow, but their simulations were restricted to a local shearing box and do not capture the global structure of the wind flow. In this paper, we aim to revisit the simulations of \citet{riols18} but in the global configuration. Our ambition is also to go one step further and produce synthetic images of the disc by computing the dust emission.  In this way, we aim to provide  predictions on the typical dust settling, ring-gap contrast and spectral index that one might observe.  \\

For that purpose, we performed global axisymmetric simulations of stratified discs with a net vertical magnetic field, including ambipolar diffusion and dust. For that, we used a customised version of the  PLUTO code, {based on PLUTO v4.3}. To simplify the problem, we considered a portion of the disc spanning from 10 to 200 AU where we could neglect the Hall effect \citep{simon15}.  The dust population is approximated as a pressure-less multi-fluid made up of different sized particles, from one hundred micrometres to one centimetre, relevant for radio-observations. 
The back reaction of the dust on the gas is taken into account, but we neglected coagulation or fragmentation processes.  To produce synthetic images of the disc, we used the radiative transfer code MCFOST. As a preliminary step, the radiative transfer calculation was not performed during the hydrodynamic steps, but was based on a fixed snapshot (the hydrodynamic calculation was post-processed  with MCFOST).  

The paper is organised as follows: In Section \ref{sec_framework}, we describe the model and review the main characteristics of non-ideal physics and dust-gas interaction.  In Section \ref{sec_numerical}, we present the numerical methods used to simulate the dust-gas dynamics in a global model. In Section \ref{sec_without_dust}, we perform numerical simulations without dust in order to characterise the properties of the disc prone to ambipolar diffusion (wind, angular momentum transport and rings). In particular, we investigate the origin of the rings and their long-term evolution. In Section \ref{sec_dust}, we explore the dynamics of the dust in the ambipolar-dominated flows.  We study, in particular, the sedimentation of the grains and show that pressure maxima associated with rings accumulate the solids into thin structures with high density contrast.  In Section \ref{sec_mcfost}, we produce  synthetic images of our disc simulation in ALMA bands by using MCFOST. Finally, in Section \ref{sec_conclusions}, we conclude and discuss  the potential implications of our work for protoplanetary disc evolution and planet formation. 
\section{Theoretical framework}
\label{sec_framework}

We model the gas and dust of a protoplanetary disc by using a global framework, in which quantities are assumed to be axisymmetric (2.5D framework).  
We note ($r$, $\theta$, $\phi$) the coordinates related to the spherical frame ($\theta$ being the co-latitude), and ($R$, $\phi$,$z$) those related to the cylindrical frame. The disc extends from an inner cylindrical radius $R_{in}=10$ AU to an outer limit $R_{out}=200$ AU.   We adopt a multi-fluid approximation in which the ionized gas and the dust interact and exchange momentum through drag forces.  

\subsection{Gas phase}
\subsubsection{Equations of motion}

The gas is coupled to a magnetic field $\mathbf{B}$ and is assumed to be inviscid and locally isothermal on cylindrical radii, its pressure $P$ and density $\rho$ related by $P=\rho c_s^2$, where $c_s(R,z)$ {is a fixed sound-speed profile  (see Section \ref{temperature} for temperature prescription).  The evolution of its density ${\rho}$, and total velocity {field} $\mathbf{v}$ follows:
\begin{equation}
\dfrac{\partial \rho}{\partial t}+\nabla\cdot \left(\rho \mathbf{v}\right)=0, 
\label{mass_eq}
\end{equation}
\begin{equation}
\frac{\partial{\rho\mathbf{v }}}{\partial{t}}+\mathbf{\nabla}\cdot (\rho \mathbf{v} \otimes \mathbf{v}) =\rho \mathbf{g}
  -\mathbf{\nabla}{P}+(\nabla\times \mathbf{B})\times\mathbf{B}+\rho\,{\bm{\gamma}_{d->g}}, 
\label{ns_eq}
\end{equation}
where the total velocity field can be decomposed into a mean Keplerian flow plus a perturbation $\mathbf{u}$: 
\begin{equation}
\mathbf{v}= R\,\Omega(R,z)+\mathbf{u}.
\end{equation}
The term $\mathbf{g}=-GM/r^2 \mathbf{e}_r$ corresponds to the gravitational field exerted by the central star, where $\mathbf{e}_r$ is the unit vector pointing in the radial direction in the spherical frame. The last term in the momentum Equation (2) contains the acceleration $\bm{\gamma}_{d->g}$ exerted by the dust drag force on a gas parcel (detailed in Section \ref{dust_equation}). {We note that the unit of time is taken as the inner $\Omega^{-1}$ and the unit of length corresponds to $R=10 AU$. The unit of mass is such that $\rho=1$ in the midplane at $R=10 AU$.}

The magnetic field $\mathbf{B}$, which appears in the Lorentz force  ${(\nabla\times \mathbf{B})\times\mathbf{B}}$, is governed by the non-ideal induction equation, 
\begin{equation}
\frac{\partial{\mathbf{B}}}{\partial{t}} =\nabla\times(\mathbf{v}\times\mathbf{B})+\mathbf{\nabla} \times \left[\eta_A (\mathbf{J} \times \mathbf{e_b}) \times \mathbf{e_b}\right].
\label{magnetic_eq} 
\end{equation}
The term $\mathbf{\nabla} \times \left[\eta_A (\mathbf{J} \times \mathbf{e_b}) \times \mathbf{e_b}\right]$  corresponds to ambipolar diffusion, with $\mathbf{J}=\mathbf{\nabla} \times \mathbf{B}$  the current density, $\mathbf{e_b}=\mathbf{B}/ \Vert \mathbf{B} \Vert$  the unit vector parallel to the field line, and $\eta_A$ the ambipolar diffusivity. To simplify the problem, we assume that the diffusion of magnetic field is only due to ion-neutral collisions (ambipolar diffusion) and is assumed to be the dominant non-ideal effect in the regions considered in this paper, that is  $R\geq 10 $ AU \citep[see justifications in][]{simon15}.  Therefore, the Hall effect and ohmic diffusion are neglected. The diffusivity $\eta_A$ is not uniform, and its vertical profile is prescribed in Section \ref{ionization_profile}.

\subsubsection{Surface density profile}

We assume a disc model with surface density following 
\begin{equation}
\label{suface_density}
\Sigma = 200\, (R/1 \,AU)^{-0.5} \text{g/cm}^{2}
.\end{equation}
This corresponds to a rather flat density profile that has been measured in some discs like HD 163296 \citep{williams16}. We note that the power-law index of the surface density inferred from observations ranges from -1.1 to 0.2 \citep{kwon15}. This choice is also motivated by theoretical considerations. Indeed, most (if not all) of the MHD models published to date assume Sigma $\propto R^{-1}$ with $H/R=cst$. This choice of density profile turns out to be peculiar
in the usual alpha disc theory \citep{shakura73}, since it predicts no accretion,
even if $\alpha$ is non-zero. This was indeed observed by \citet{bethune17} and \citet{bai17}. We therefore choose a shallower
disc profile, so that even if no wind is present, turbulence alone could
drive a measurable accretion rate (see Section \ref{angular_momentum_budget}).

\subsubsection{Ambipolar  profile}
\label{ionization_profile}

We consider here that the only charged particles are the ions and electrons. The ambipolar diffusivity $\eta_A$ is given by:
\begin{equation*}
\eta_A=\dfrac{\mathbf{B}^2}{\gamma_i \rho_n  \rho_i}
\end{equation*}
\citep{wardle07}, where $\gamma_i=\langle \sigma v \rangle_i /(m_n+m_i), $ and $\langle \sigma v \rangle_i=1.3\times 10^{-9} \text{cm}^3 \text{s}^{-1}$ is the ion-neutral collision rate. $\rho_n$ and $\rho_i$ are, respectively, the neutral and ion mass density; $m_n$ and  $m_i$ their individual mass. By introducing the Alfv\'en speed $v_A=B/\sqrt{\rho}$, we also define the dimensionless ambipolar Els\"{a}sser number
\begin{equation*}
\text{Am}=\dfrac{v_A^2}{\Omega \eta_A} = \dfrac{\gamma_i \rho_i}{\Omega}.
\end{equation*}
To calculate $\rho_i$,  one generally needs to take into account the various ionisation sources (X rays, cosmic rays, radioactive decay and FUV from the star) as well as the dissociative recombination mechanisms. This requires the computation of the complex chemistry occurring in the gas phase and on grain surfaces. To simplify the problem as much as possible, we use a crude model that captures the essential physics of disc ionisation, which is the effect of FUV in the disc corona.  In this model,  Am is constant and of order unity in the midplane, and it increases abruptly up to a certain height $z_{io}$.  The height $z_{io}$ corresponds to the base of the ionisation layer, above which the FUV can penetrate. If we note $\Sigma_i(z)$ as the horizontally averaged column density, integrated from the vertical boundary, we assume that FUV are blocked for $\Sigma_i(z)>\Sigma_{ic}=0.0005$ g/cm$^{2}$.  This corresponds to $z_{io}\simeq 3.5 H$, where $H$ is the gas disc scale height. To avoid any discontinuity at $z=z_{io}$, we use a smooth function to make the transition between the midplane regions and the ionised layer, so that the Ambipolar Elsasser number is
\begin{equation}
\label{eq_Am}
\text{Am} = \text{Am}_{0}\, \exp\left[ \left( \dfrac{\Sigma_{ic}}{\Sigma^+}\right)^2+\left( \dfrac{\Sigma_{ic}}{\Sigma^-}\right)^2\right],
\end{equation}
where $\text{Am}_{0}$  is the  constant midplane value, $\Sigma^+$ is the column density integrated from the top boundary, and  $\Sigma^-$ the column density integrated from the bottom boundary. {This diffusivity profile is designed to mimic the conductivity obtained from complete thermochemical models of protoplanetary discs including dust grains, with $\text{Am}_{0}\simeq 1$.} \citep{thi18}. We note that to convert numerical values of $\Sigma_i$  into g/cm$^{2}$, we used the disc model of Eq.~(\ref{suface_density}).

\subsubsection{Temperature and hot corona}
\label{temperature}
Complex thermo-chemical models of protoplanetary discs  \citep{kamp04,thi18} suggest that there is a sharp transition in temperature between the disc midplane and its ionised corona. The warm and optically thin corona can reach temperatures of a few thousand Kelvin, while the disc temperature remains below 100 K. The corresponding ratio of corona to midplane sound speed ranges typically from three to 10. To reproduce this temperature structure, we force the gas to be warmer above a certain height $z_{io}\simeq 3.5 H$. This is chosen to be the same altitude as the one at which we impose a jump in ambipolar Elssaser number (Am). We  take  a  constant  corona-to-midplane temperature  ratio  at  all  radii equal to six. Again to avoid any discontinuity, we use a smooth function similar to that of Eq.~\ref{eq_Am} to make the transition between the two regions. 

\begin{equation}
T=T_{mid}(R)+\left[T_{cor}(R)-T_{mid}(R)\right]*[1-f(z,R)]
\end{equation}
with 
\begin{equation}
f(z,R)=\exp\left[-\left( \dfrac{\Sigma_{ic}}{\Sigma^+}\right)^2-\left( \dfrac{\Sigma_{ic}}{\Sigma^-}\right)^2\right].
\end{equation}
We note that inside the two distinct regions (disc and corona), {we force the flow to maintain a constant temperature on cylindrical radii. For that, we evolve the energy equation with rapid cooling to keep the gas close to being locally isothermal (with a cooling time proportional to $\Omega^{-1}$).}

\subsection{Dust phase} 

\subsubsection{Equations of motion} 
\label{dust_equation}
The dust is composed of a mixture of different species,  characterising different grain sizes. Each species, labelled by the subscript $k$, is described by a pressureless fluid,  with a given density $\rho_{d_k}$ and velocity $\mathbf{v}_{d_k}$. We assume in this paper that dust grains remain electrically neutral so that they do not feel the magnetic field. The equations of motion for each species are: 
\begin{equation}
\dfrac{\partial \rho_{d_k}}{\partial t}+\nabla\cdot \left(\rho_{d_k} \mathbf{v_{d_k}}\right)=0,
\label{eq_mass_dust}
\end{equation}
\begin{equation}
\dfrac{\partial \rho_{d_k} \mathbf{v_{d_k}}}{\partial{t}}+\mathbf{\nabla}\cdot (\rho_{d_k} \mathbf{v_{d_k}} \otimes \mathbf{v_{d_k}}) =\rho_{d_k} \,(\mathbf{g}+ \bm{\gamma}_{g->{d_k}}),
\label{ns_eq_dust}
\end{equation}
with $\bm{\gamma}_{g->{d_k}}$ the drag acceleration induced by the gas on a dust particle.  We obtain, by conservation of total angular momentum:
\begin{equation*}
\bm{\gamma}_{d->g}= -\dfrac{1}{\rho} \sum_k \rho_{d_k} \,\bm{\gamma}_{g->{d_k}}.
\end{equation*}
The acceleration associated with the drag and acting on a single particle is given by:
\begin{equation*}
 \bm{\gamma}_{g->{d_k}}=\dfrac{1}{\tau_{s_k}}  (\mathbf{v}-\mathbf{v_{d_k}}),
\end{equation*}
where $\tau_{s_k}$ is the stopping time that  corresponds to the timescale on which frictional drag causes a significant change  in the momentum of the dust grain. This is a direct  measure of the coupling between dust particles and gas. To evaluate $\tau_{s_k}$, we consider in this study that dust particles are spherical and small enough\footnote{{We note that in principle, dust particles could be porous and have a more complex geometry. This would not change the dynamics described here, but only the relation between the Stokes number and the physical size of the particles (\ref{eq_stoptime}).}} to be in the Epstein regime \citep{weiden77}. For a spherical particle of size $a_k$ and internal density $\rho_s$ (which should not be confused with the fluid density $\rho_d$), the stopping time is:
\begin{equation}
\label{eq_stoptime}
\tau_{s_k}  = \dfrac{\rho_s {a_k}}{\rho c_s}.
\end{equation}
A useful dimensionless quantity to parametrise the coupling between dust and gas is the Stokes number: 
\begin{equation*}
\text{St}=\Omega \tau_{s_k}.
\end{equation*}

\subsubsection{Converting  particle size to Stokes number}
\label{conversion}
{In this paper, we keep the grain size $a_k$ of each dust component fixed in space and time. To make an easier comparison with the literature, we also label each dust component by its Stokes number measured for $\rho=\rho_0=1$, $c_s=c_{s_0}=0.05$ and $\Omega=\Omega_0=1$,  which we label $\text{St}_0$ and refer to as the 'reference Stokes number'. We note, however, that the local Stokes number (which controls the dynamics) depends on the position in the disc for one species.}  Therefore, we need to find a correspondence between the grain size and the Stokes number. For that, we assume that in the limit of small magnetisation, the disc is in hydrostatic equilibrium and its column (or surface) density is  $\Sigma = \rho_0  H \sqrt{2\pi,}$ where $\rho_0$ is the midplane density and $H= c_s/\Omega$ is the disc scale height. Thus, combining these different relations, we obtain:
\begin{equation*}
\text{St}=a_k \left( \dfrac{\rho_s \sqrt{2\pi}}{\Sigma}\right)
\end{equation*}
in the midplane.
To estimate the Stokes number as a function of grain size, we  assume that the disc surface density follows Eq.~\ref{suface_density}. 
By considering $\rho_s=2.5$ g/cm$^{-3}$,  we obtain the following conversion in the midplane:
\begin{equation}
\label{eq_conversion}
\text{St} \simeq \text{St}_0\, \left(\dfrac{R}{ \text{10 AU}}\right)^{1/2}   \quad \text{with} \quad  \text{St}_0 = 0.1 \,\,  \left(\dfrac{a_k}{\text{1 cm}}\right) .
\end{equation}
\subsection{Initial equilibrium}

For the initial hydrostatic disc equilibrium,  we use the same prescription as \citet{zhu18}: 
\begin{equation}
\label{eq_rho_init}
\rho = \rho_0 \left(\dfrac{R}{R_0}\right)^{-3/2}\, \exp\left[\dfrac{GM}{c_s^2} \left( \dfrac{1}{\sqrt{R^2+z^2}}-\dfrac{1}{R}\right)\right]
,\end{equation}
\begin{equation}
\label{eq_vphi_init}
v_\phi = \left(\dfrac{GM}{R}\right)^{1/2} \left(\dfrac{R}{\sqrt{R^2+z^2}}-2.5 c_s^2 R\right)^{1/2}  
,\end{equation}
where $c_s=\sqrt{P/\rho}$ is the sound speed that is constant on the cylindrical radius:
\begin{equation}
\label{eq_cs_init}
c_s= {c_{s_0}} \left(\dfrac{R}{R_0}\right)^{-1/2}
\end{equation}

The subscript `0' denotes quantities  at the inner radius in the midplane. Taking Eq.\ref{eq_rho_init} and developing at first order in $z/R$, we find that the disc is in a classical hydrostatic equilibrium  with  density  varying on  a  characteristic  vertical  scale  $H=c_s/\Omega$.
Considering the fact that $\Omega \propto R^{-3/2}$ and $c_s$ given by Eq.~\ref{eq_cs_init}, the ratio $H/R$ is constant with radius and is given  by $c_{s_0}/v_{K_0}$. 
Protoplanetary  discs  are  geometrically  thin, typically with $H/R$ between  0.03 and 0.2 (Bitsch et al. 2013). In this paper, we thus choose $H/R=c_{s_0}/v_{K_0} =0.05$.  A constant $H/R$ is probably not the most encountered configuration in observed systems, although it is in line with some objects like HK Tau B or Oph163131. 

We note that our initialisation is problematic  at the pole, since the density, azimuthal velocity and sound speed become infinite at that location. To avoid this, we use $R=\max(R,R_{min})$ in the above equations with $R_{min}=10$ AU. The magnetic field is initialised in a way similar to that in \citet{zhu18}. Initially, there is a pure vertical field component, which has the following dependence on R:
\begin{equation*}
B_z = 
\begin{cases}
B_0 \left(\dfrac{R_{min}}{R_0}\right)^m \quad \text{if} \, R<R_{min}\\\\
B_0 \left(\dfrac{R}{R_0}\right)^m \quad \text{if}\,  R>R_{min}
\end{cases}
\end{equation*}
with $m=-2.5/2$ and $R_{min}=R_0=10$ AU. Again the prescription at $R<R_{min}$ is to avoid the divergence of the vertical field at the pole.  With such dependence, it is possible to define a constant magnetisation or plasma $\beta$ parameter in the midplane for $R>R_{min}$ : 
\begin{equation*}
\beta=\dfrac{2\rho c_s^2}{B_z^2} = \dfrac{2\rho_0 c_{s_0}^2}{B_0^2}
\end{equation*}

\subsection{Averages}

We define a vertical integral of a quantity $Q$ between angles $\theta_-$ and $\theta_+$ as 
\begin{equation}
\label{theta_average}
\overline{Q}=\int_{\theta_-}^{\theta_+} r \sin{\theta} Q d\theta
,\end{equation}
and, respectively, a radial integral and temporal average: 
\begin{equation}
\label{radial_average}
\langle{Q}\rangle_R =\int_{r_{in}}^{r_{out}}  Q dr \quad \text{and} \quad \langle{Q}\rangle = \dfrac{1}{T_f-T_i}\int_{Ti}^{T_f}  Q dt
.\end{equation}

\subsection{Conservation of mass and angular momentum in spherical coordinates}

Using the average defined in Eq.\ref{theta_average}, the mass conservation equation then reads, in spherical coordinates:
\begin{equation}
\dfrac{\partial \Sigma}{\partial t} + \dfrac{1}{r} \dfrac{\partial}{\partial r} r \overline{\rho u_r} +\dot{\sigma}_w=0
,\end{equation}
where $\Sigma=\overline{\rho}$ is the gas surface density defined on spherical geometry and $\dot{\sigma}_w= \left[\rho u_\theta \sin{\theta}\right]_{\theta_-}^{\theta_+}>0$ is the mass loss rate measured between angles  $\theta_-$ and  $\theta_+,$ {which mark the location of the disc surface.} If we note the vertically integrated radial stress between $\theta_-$ and  $\theta_+$:
\begin{equation}
\overline{\mathcal{T}}_{r\phi}= \overline{(\rho u_\phi u_r - B_\phi B_r) \sin{\theta}}
,\end{equation}
and the  vertical stress difference between the two angles: 
\begin{equation}
{\mathcal{T}}_{\theta\phi}= \left[(\rho u_\phi u_\theta - B_\phi B_\theta) r\sin^2{\theta}\right]_{\theta_-}^{\theta_+},
\end{equation}
the conservation of angular momentum in spherical coordinates reads: 
\begin{equation}
\label{angular_mometum1}
\overline{\rho u_r} \dfrac{\partial}{\partial r} \left[\Omega(r) r^2 \right] + \dfrac{1}{r} \dfrac{\partial}{\partial r} r^2 \overline{\mathcal{T}}_{r\phi} +{\mathcal{T}}_{\theta\phi} =0 
,\end{equation}
with $\Omega(r)=\sqrt{GM/r^3}$, where we assumed that $u \ll \Omega R,$ 
which allows us to neglect the time derivative term.
We define, respectively, the accretion rate within the disc and the mass outflow rate  as: 
\begin{equation}
\dot{M}_a = -2\pi r\overline{\rho u_r} \quad \text{and} \quad \dot{M}_w =\int_{r_{in}}^r 2\pi r \dot{\sigma}_w dr
.\end{equation}
Multiplying Eq.~(\ref{angular_mometum1}) by $r$ and integrating in radius,  we obtain the following conservation equation: 
\begin{equation}
\label{angular_mometum2}
\left\langle r\dfrac{\dot{M}_a  \Omega}{4 \pi} \right\rangle_R=  \left[r^2 \overline{\mathcal{T}}_{r\phi}\right]_{r_{in}}^{r_{out}} +\left\langle{r\mathcal{T}}_{\theta\phi}\right\rangle_R
.\end{equation}
Accretion onto the star can be either due to a radial transport of angular momentum, via the radial stress $\overline{\mathcal{T}}_{r\phi}$ , or due to angular momentum extracted by a wind through the disc surface, via the vertical stress ${\mathcal{T}}_{\theta\phi}$.  
An important quantity that characterises the radial transport efficiency is the parameter $\alpha$ (or Shakura \& Sunyaev parameter)  which is: 
\begin{equation}
\label{eq_stress}
\alpha =  \dfrac{\overline{\mathcal{T}}_{r\phi}}{\overline{P}}=\dfrac{\overline{\mathcal{R}}_{r\phi}}{\overline{P}} + \dfrac{\overline{\mathcal{M}}_{r\phi}}{\overline{P}} = \alpha_R+\alpha_M
,\end{equation}
where $\overline{\mathcal{R}}_{r\phi}=\overline{\rho u_\phi u_r \sin{\theta}}$ is the Reynolds stress and $\overline{\mathcal{M}}_{r\phi}=\overline{-B_\phi B_r\sin{\theta}}$ is the Maxwell stress. The coefficient $\alpha$ can be decomposed into a sum of a laminar part: 

\begin{equation}
\label{eq_laminar_stress}
\alpha_L = \dfrac{ \overline{ \left(\langle \rho \rangle  \,  \langle u_\phi \rangle \langle u_r \rangle  -\langle{ B}_\phi \rangle \langle {B}_r\rangle\right) \sin{\theta}}}{ \overline{\langle P \rangle} }
,\end{equation}
plus a turbulent part: $\alpha_\nu = \alpha - \alpha_L$.

\section{Numerical setup}
\label{sec_numerical}

\begin{figure*}
\centering
\includegraphics[width=0.49\textwidth]{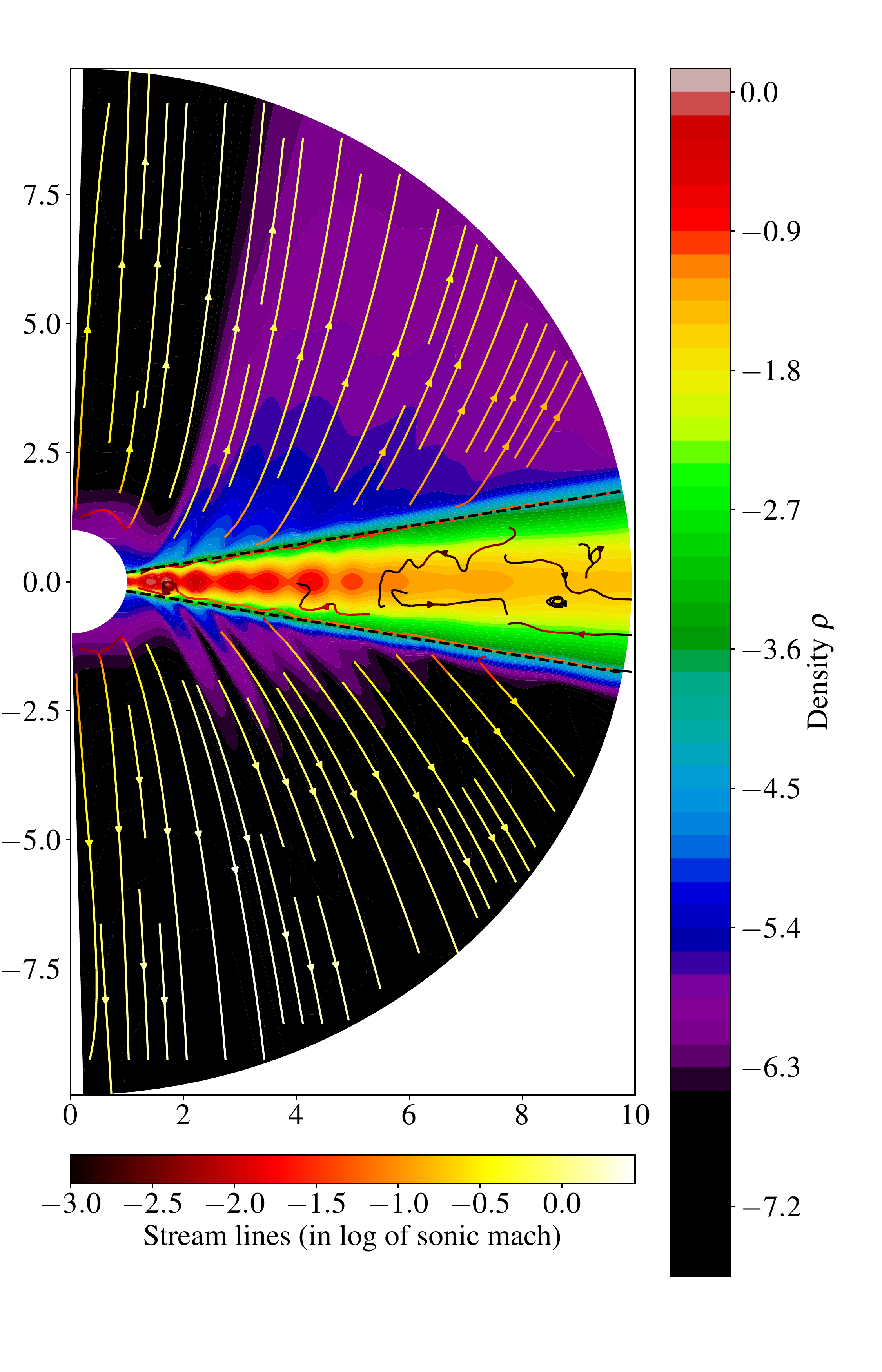}
\includegraphics[width=0.49\textwidth]{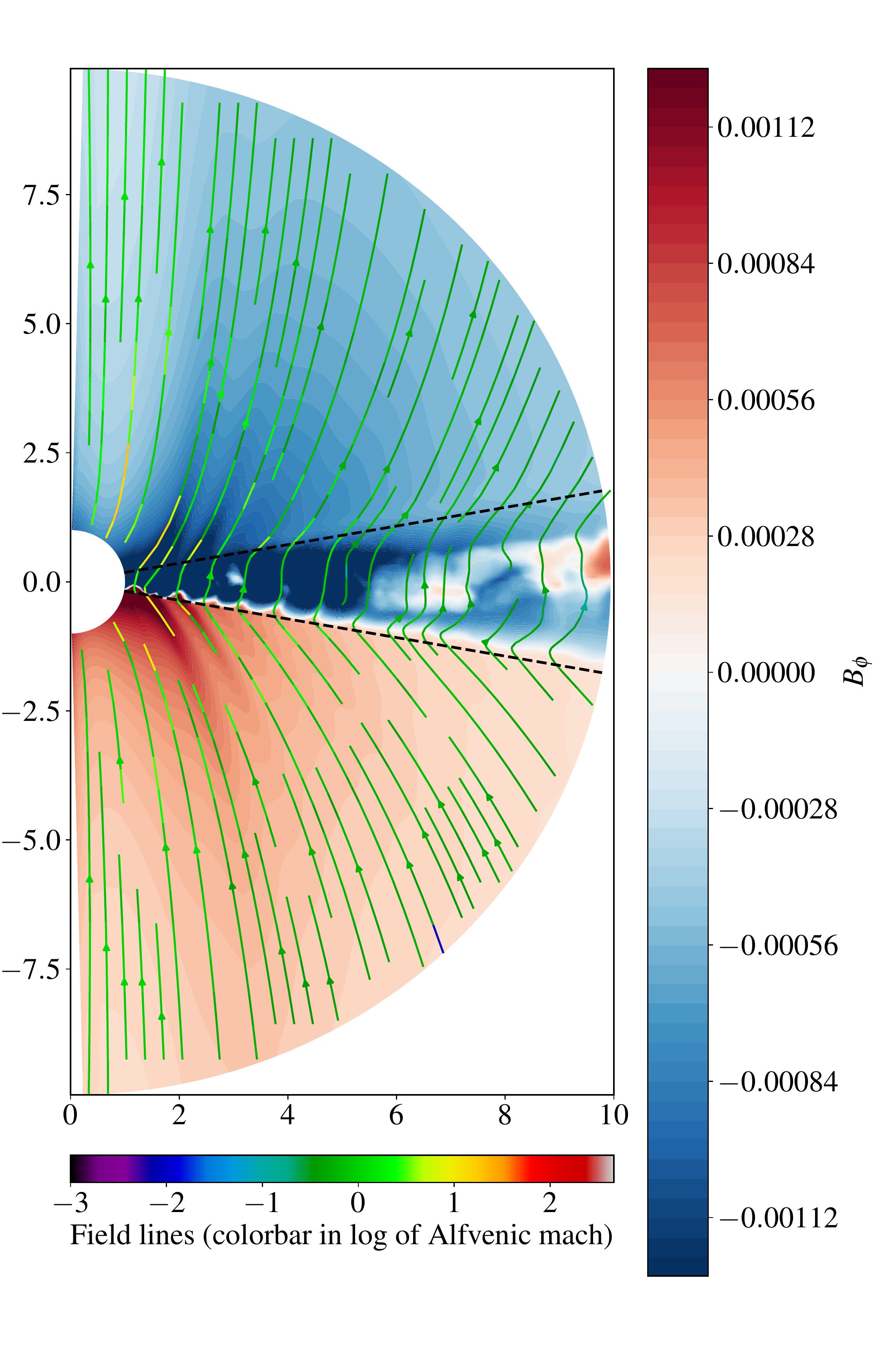}
 \caption{Left: Average gas density and streamlines in the poloidal plane. Right: Average toroidal field and poloidal field lines. All quantities are averaged between $t=250 T_0$ and $t=1000 T_0$. }
\label{fig_Blines_hot}
\end{figure*} 

\begin{figure}
\centering
\includegraphics[width=\columnwidth]{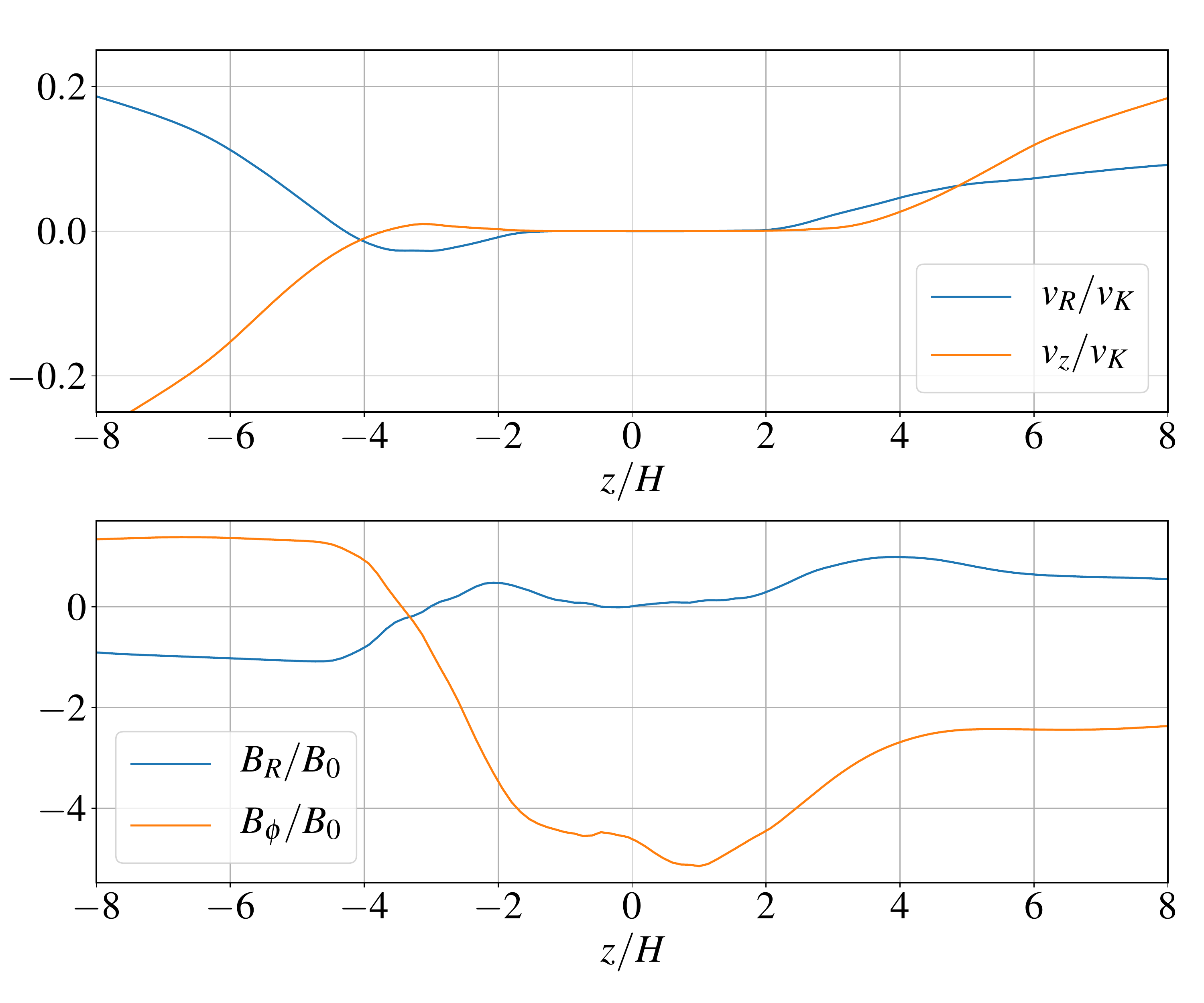}
 \caption{Top: Vertical profiles of the radial and vertical velocity. Bottom:  Vertical profiles of the radial and toroidal magnetic fields. Quantities are averaged between $R=25 AU$ and $R=50 AU$  and in time between $t=250 T_0$  and $t=1000 T_0$. }
\label{fig_zprofile}
\end{figure} 

\begin{figure}
\centering
\includegraphics[width=\columnwidth]{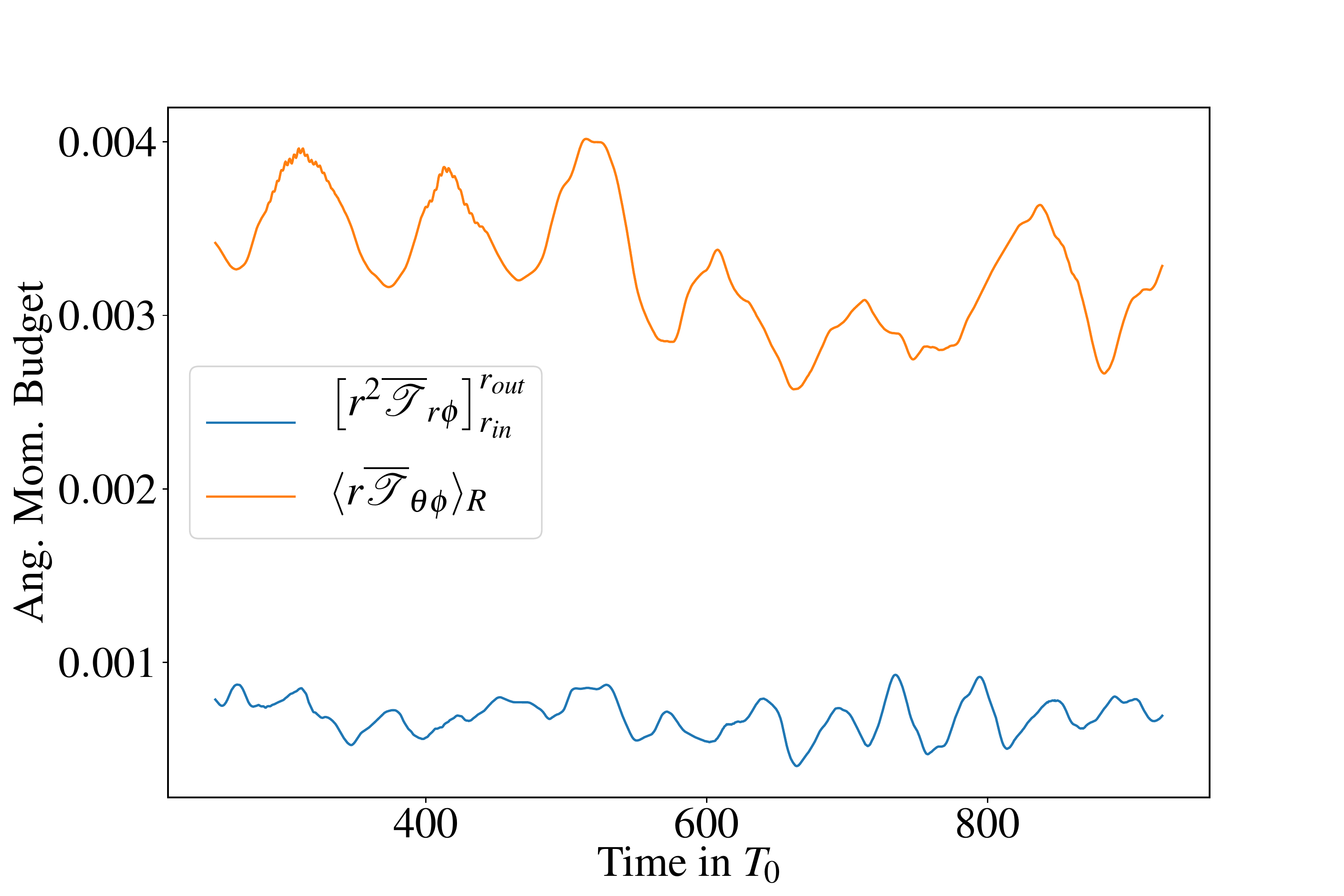}
 \caption{Time evolution of the radial (blue) and vertical  (orange) torques, integrated radially and vertically, appearing in the right hand side of the angular momentum equation (Eq.~\ref{angular_mometum2})}
\label{fig_angmom}
\end{figure} 

\begin{figure}
\centering
\includegraphics[width=\columnwidth]{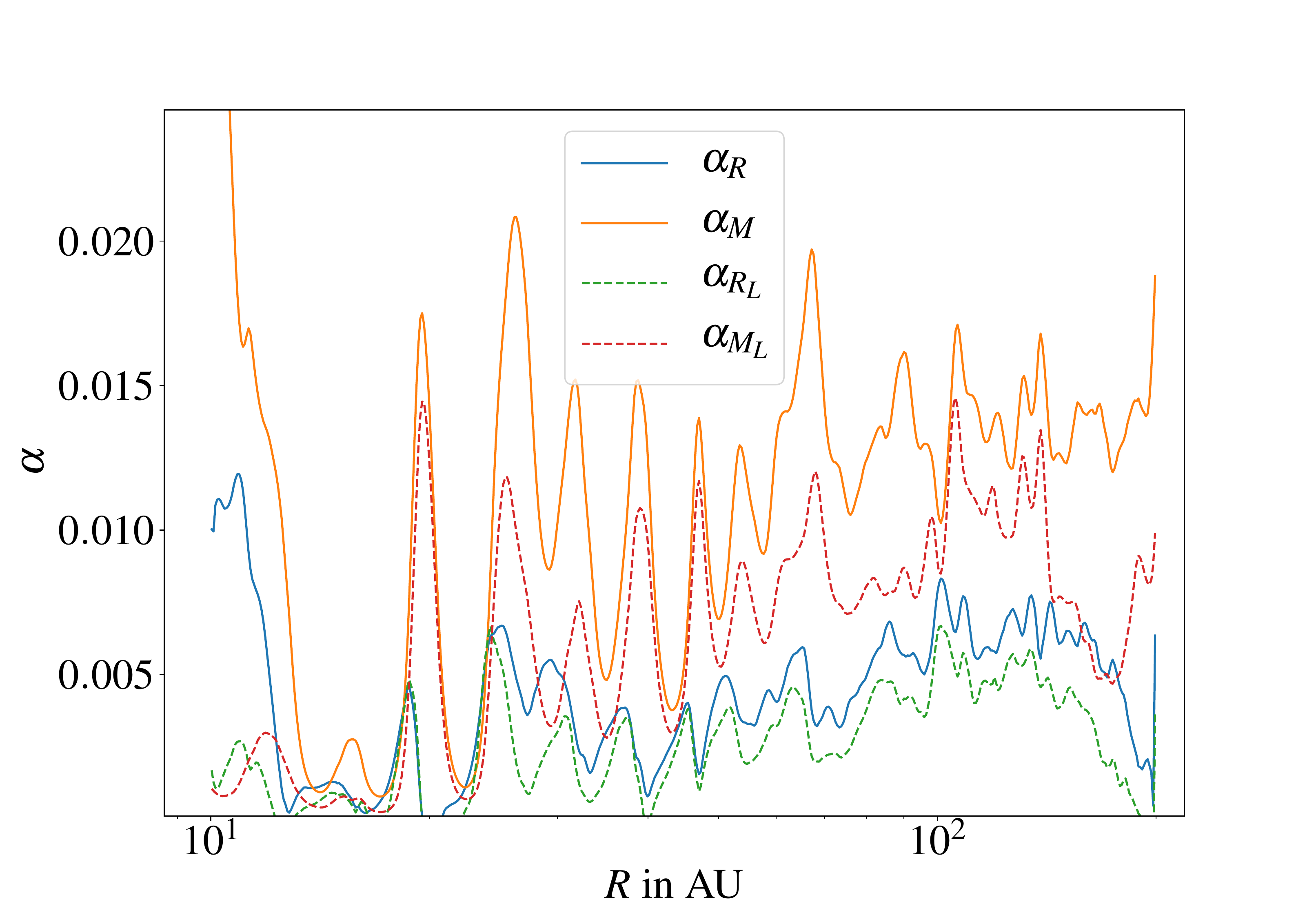}
 \caption{Profile in $R$ of the radial transport coefficient $\alpha_R$  and  $\alpha_M$ (defined in Eq.~\ref{eq_stress}) averaged in time between 250 and 1000 orbits. The dashed line shows the laminar contribution to these coefficients.}
\label{fig_alpha}
\end{figure}

\subsection{Numerical code}
To simulate the dynamics of the gas and dust,  we use a modified version of the Godunov-based PLUTO code \citep{mignone07}.   This code solves the approximate Riemann problem at each cell interface and is written so that it ensures conservation of mass and angular momentum. Inter-cell fluxes are computed with a HLLD solver for the gas and an HLL solver for the dust. We use a piecewise linear space reconstruction to estimate the Godunov fluxes at cell interfaces, with the monotonised central limiter.  Physical quantities are  evolved in time with  a second  order explicit Runge-Kutta scheme. Parabolic terms, including ambipolar diffusion, are also solved explicitly and introduce a strong constraint on the CFL condition.  We note that we do not use the FARGO orbital advection scheme, as the CFL is limited by the parabolic terms due to ambipolar diffusion. 

\subsection{Grid and resolution}

In radius, we use a logarithmic grid with 512 points, spanning from 10 AU to 200 AU. The increment $\delta r$ is then proportional to the local radius $r$.  In $\theta$, the  grid is split into three different regions. The first one is the midplane region from $\theta=1.28$ to $\theta=1.86$ (corrsponding to $\pm 16^\circ$ on either side of the midplane), which has 96 uniformly distributed cells. This corresponds to a vertical segment spanning $\pm 5$ scale height, with a resolution of nine points per $H$. The second and third regions are symmetric  around the mid-plane and  correspond, respectively, to $\theta=[0, 1.28]$ and $\theta=[1.86, \pi]$. In each case, the grid is stretched and contain 64 points. Our grid resolution in $r$ and $\theta$  is enough to capture most of the dynamics and the structures encountered in ambipolar-dominated discs. 

\subsection{Boundary and internal conditions}

At the inner radius, we impose outflow boundary conditions for all the quantities except for the azimuthal velocity, on which we imposed the initial keplerian velocity given in Eq.\ref{eq_vphi_init}. We also make sure that $v_r<0$ to prevent any matter from entering the disc.   At the outer radius, we apply similar outflow condition, but we do not force the keplerian equilibrium. 
In the $\theta$ direction, we implement a specific boundary condition allowing our simulations to extend all the way to the pole (from $\theta=0$ to $\theta=\pi$).  For that boundary, we use a similar implementation  to \citet{zhu18} (see their Appendix). 

To avoid small time steps,  we limit the Alfv\'en velocity in the whole domain to eight times the Keplerian velocity at the inner disc radius {: $v_A/v_{K_0} <8$}. This is done by artificially increasing the density in the cells where this condition is violated. In general, this affects the cells in the disc's corona rather than the cells in the bulk of the disc. 
In addition to that and in order to stabilise the code, we also introduce a floor in the density, independent of the floor in Alfv\'en velocity. The condition reads: 
\begin{equation*}
\rho>\rho_0\max{\left(10^{-8},  10^{-6}\dfrac{ (R/R_0)^{-3/2}}{(z/R_0)^2+1.2 (H/R)^2 R/R_0}\right)}.
\end{equation*}
We impose a density floor for the dust as well so that $\rho_d$ never decrease below $10^{-14}$ in code units.
At each time step, we also relax the temperature towards its initial value over a characteristic timescale of $0.1 \Omega^{-1}$, so that both the disc and the corona remain locally isothermal during the course of the simulation (each of them relaxing toward a different temperature or sound speed). We note that this artificial relaxation is similar to a rapid cooling of the flow, which prevents vertical shear instability. Finally to avoid spurious and fast accretion of the field lines near the inner boundary, we relax the poloidal velocity toward zero for $R<15 AU$ within a characteristic timescale equal to $0.5\Omega^{-1.}$

\section{Ambipolar simulation without dust}
\label{sec_without_dust}

\subsection{General properties of the outflow}
\label{general_properties}

 In this section, we study a global axisymmetric and pure gaseous simulation with $\beta=10^3$ and $ \text{Am}_0=1$. This simulation is integrated over 1000 $T_0$, where $T_0$ is the orbital period at the inner radius. In Fig.~\ref{fig_Blines_hot} (left), we show the time-averaged density and streamlines in the inner region between $R=10$ and $R=100$ AU.  The upper (respectively lower) base of the disc is defined by the surface at angle  $\theta_-=\pi/2- \theta_d$  (respectively $\theta_+=\pi/2+ \theta_d$ ) with $\theta_d\simeq10 ^{\circ}$.  They are  delimited by the dashed black lines in Fig.~\ref{fig_Blines_hot} and correspond to the frontiers at $z\simeq 3.5H$ between the  ambipolar-dominated region and the fully ionised atmosphere. This angle corresponds also to the jump in temperature, meaning the separation between the cold disc and the hot corona. We clearly see that a wind is emitted and accelerated from the base of the disc at $\theta\simeq \pi/2\pm \theta_d$, with supersonic speed higher in the atmosphere. The total mass outflow rate, averaged in time and integrated  between the inner and outer radius is $\dot{M}_w(r_{out})\simeq 6\times 10^{-8} M_\odot$/yrs. One particularity is that the wind is highly dissymmetric  around the midplane, with a more massive outflow in the northern hemisphere ($\dot{\sigma}_w$ {being} four times larger than in the southern hemisphere). 
This is similar to the configuration depicted in Figs.~21 and 22  of \citet{bethune17}. Figure~\ref{fig_zprofile} (top) shows the vertical profile of the radial and vertical velocities, integrated in time and in radius between $R=25$ and $R=50 AU$. We find that the outflow in the northern hemisphere, which carries more mass, appears slower than the wind in the southern hemisphere. 
Figure~\ref{fig_Blines_hot} (right panel) shows the topology of the averaged magnetic field threading the disc, while Fig.~\ref{fig_zprofile} (bottom) displays the vertical profile of the radial and toroidal components. In the inner regions  ($\lesssim 50 AU$), the field lines cross the midplane with some inclination and bend outside in the southern hemisphere, around $z \simeq 3.5H$.  A remarkable  feature is that the toroidal magnetic field keeps its polarity and is almost constant within the midplane. There is a transition between a negative toroidal field and a positive field at $R\simeq 60AU$. Another important feature that we discuss later in Section \ref{rings_origin} is the formation of rings, which are visible in Fig.~\ref{fig_Blines_hot} (top) and appear after 200 inner orbits in the simulation.

\begin{figure}
\centering
\includegraphics[width=\columnwidth]{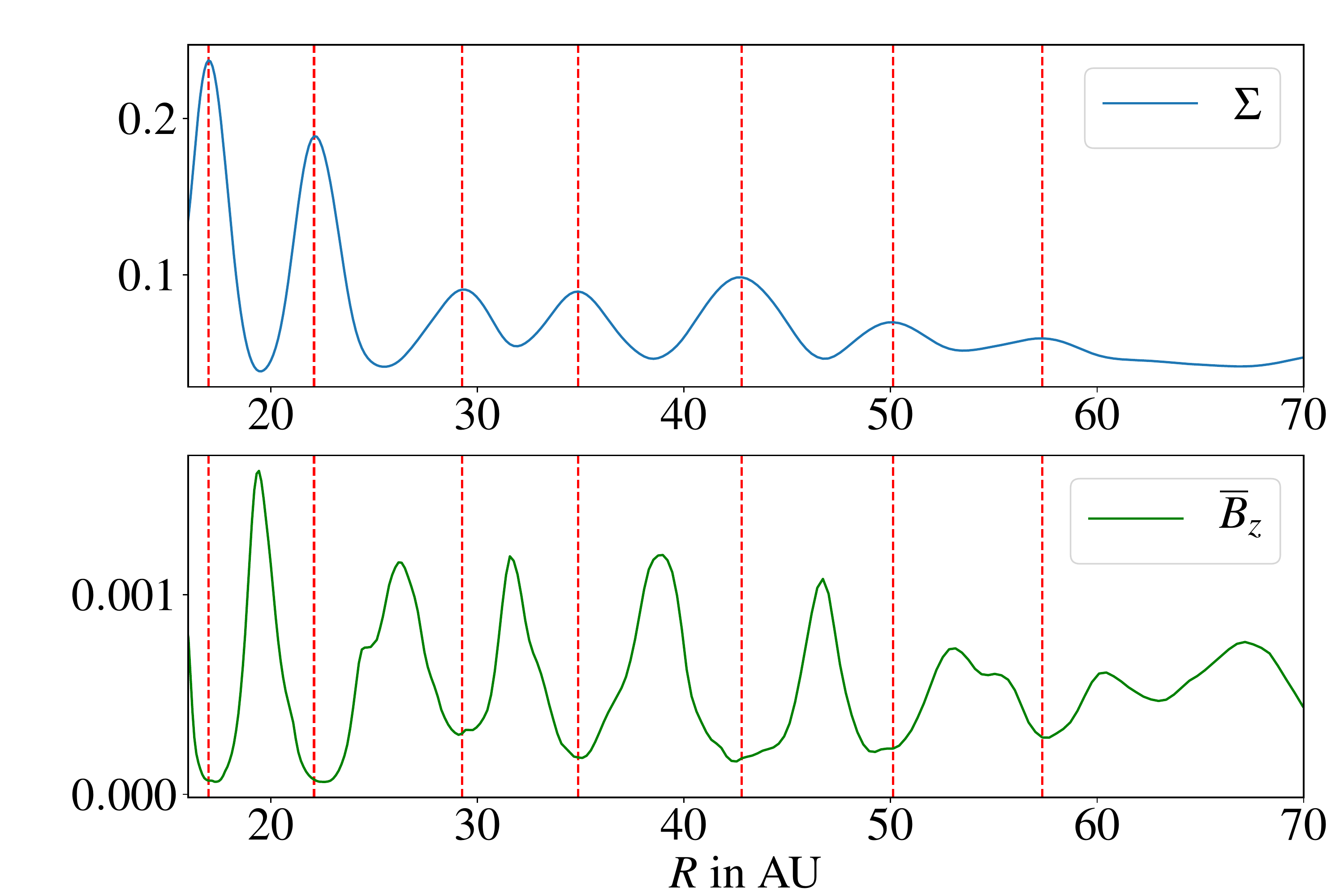}
 \caption{Surface density $\Sigma$ defined as $\overline{\rho}$ (top) and vertically-integrated magnetic field $\overline{B}_z$ (bottom) as a function of radius R. Quantities are averaged in time between 250  and 1000 orbits, and vertically between $\theta=\theta_-$ and $\theta=\theta_+$ $(z\simeq \pm 3.5H)$. The dashed red lines are the location of the density maxima. }
\label{fig_rhobz}
\end{figure} 

\begin{figure}
\centering
\includegraphics[width=\columnwidth]{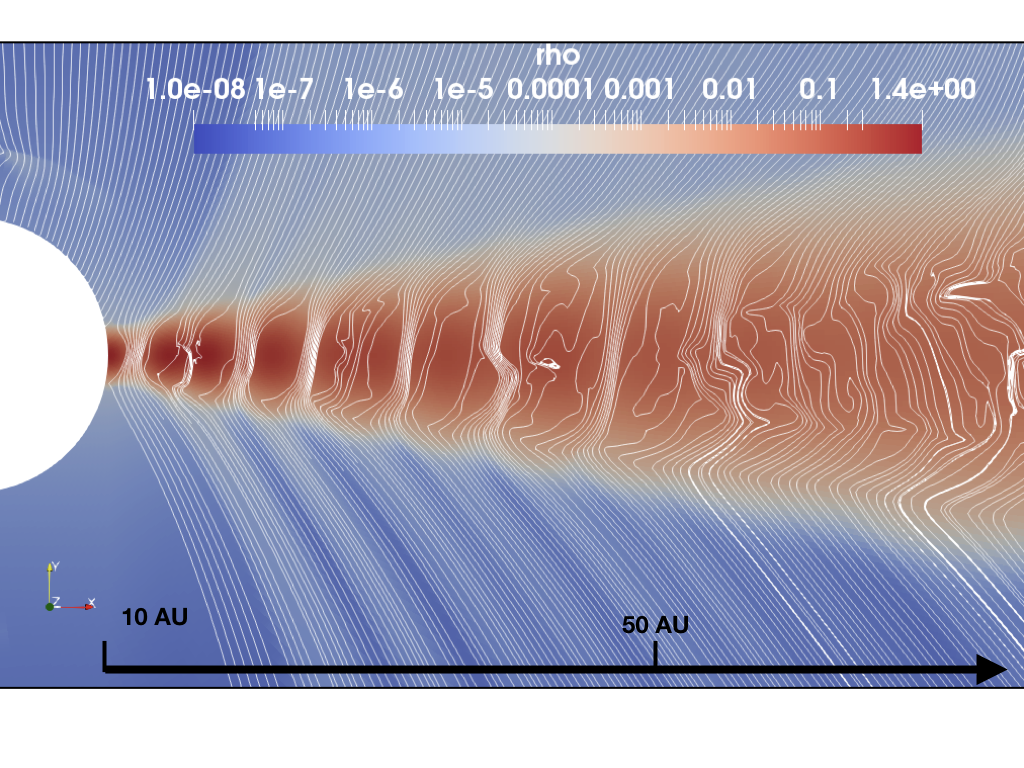}
 \caption{Snapshot of the simulation for $\beta=10^3$ at $t=350 T_0$, zoomed in the inner part of the disc. The colour map represents the density, while the white lines are the poloidal magnetic field lines.}
\label{fig_zonal_flow}
\end{figure} 

\begin{figure*}
\centering
\includegraphics[width=0.9\textwidth]{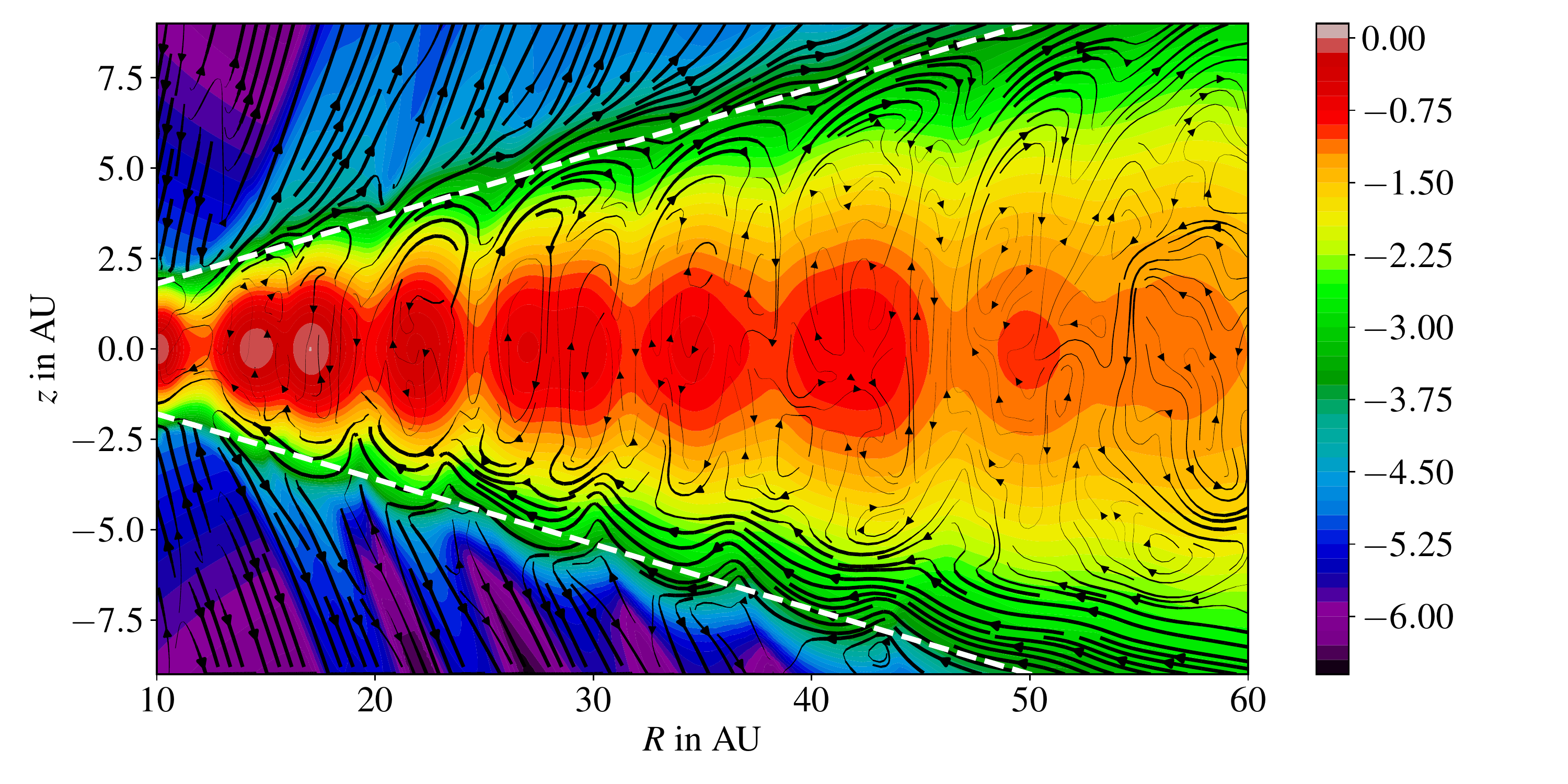}
 \caption{Snapshot of the simulation for $\beta=10^3$ at $t=350 T_0$, zoomed in on the inner part of the disc. The colour map represents the density, while the black lines are the poloidal streamlines. The dashed white lines delimit the $z =\pm 3.5H$ surface.}
\label{fig_streamlines}
\end{figure*} 

\begin{figure}
\centering
\includegraphics[width=\columnwidth]{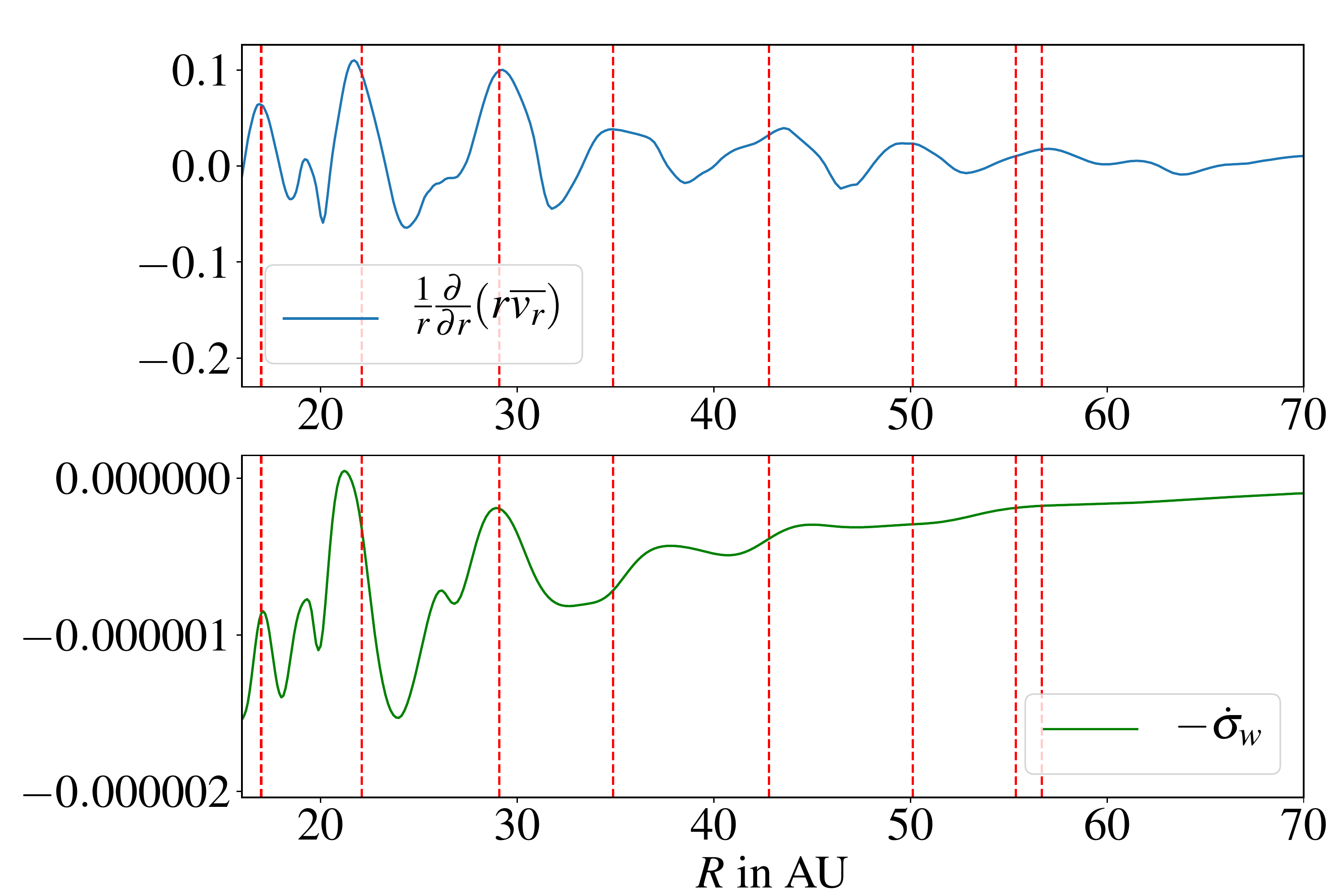}
 \caption{Top: Divergence of the radial flow integrated vertically between $\theta=\theta_-$ and $\theta=\theta_+$ $(z\simeq \pm 3.5H)$ as a function of radius $R$. Bottom: mass loss rate $\dot{\sigma}_w$  as a function of radius $R$ (computed at $z\simeq \pm 5H$). Quantities are averaged in time between 400 and 650 orbits. The dashed red lines are at the location of the density maxima. }
\label{fig_massbudget}
\end{figure} 

\begin{figure*}
\centering
\includegraphics[width=\textwidth]{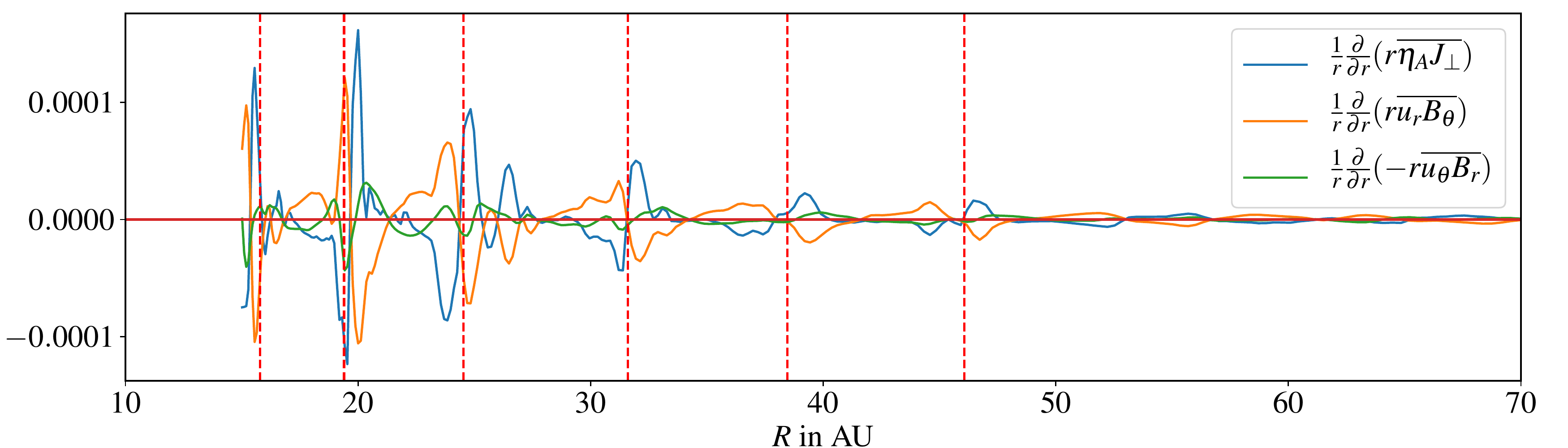}
 \caption{Radial profiles of the ideal and non-ideal terms in the induction equation for $B_\theta\simeq -B_z$ (Eq.~\ref{eq_induction_Bz}), integrated vertically between $\theta=\theta_-$ and $\theta=\theta_+$ $(z\simeq \pm 3.5H)$ and averaged in time between $t=300$ and $t=500 T_0$. The red dashed lines correspond to the local maxima of $B_z$ (averaged vertically and in the same time window).}
\label{fig_induction}
\end{figure*} 

\subsection{Angular momentum budget}
\label{angular_momentum_budget}
We find that the accretion rate in the disc, averaged in time, follows $\dot{M}_a=9 \times 10^{-8}(R/10\, \text{AU})^{0.6} M_\odot$/yrs. This is roughly five times larger than the cumulative mass outflow rate at 100 AU. {The dependence on $R$ of the mass accretion rate is expected in this situation as a result of mass conservation, and is similar to the one found in previous models \citep{bethune17}. It is a known signature of weakly magnetised magneto-thermal winds.} 
To characterise the origin of this accretion flow, in Fig.~\ref{fig_angmom} we show the two different terms appearing in the angular momentum budget (Eq.~\ref{angular_mometum2}). 
We find that the vertical transport of angular momentum, associated with the wind stress  ${\mathcal{T}}_{\theta\phi}$ is the main driver of accretion. This result is similar to that of \citet{bethune17}, although in our case the radial transport is not completely negligible, {probably because the surface density profile we use here is different from that of \cite{bethune17}}.  We checked that the vertical  flux of angular momentum is dominated by its Maxwell component $-B_zB_\phi$. Because the $B_z$ and $B_\phi$ field keep the same sign within the disc, the vertical  flux  of  angular  momentum $\mathcal{M}_{z\phi}=-B_zB_\phi$ is  unidirectional in the disc bulk. The  angular momentum is extracted from the disc southern surface where the gradient $\partial_z\mathcal{M}_{z\phi}$ is maximum (i.e. where $B_\phi$ changes its sign) and is transported  towards the northern hemisphere.

 In Fig.~\ref{fig_alpha}, we also show the radial profile of the vertically averaged radial transport coefficients $\alpha_R$ and $\alpha_M$.  This is averaged between 250 and 1000 $T_0$. The radial transport of angular momentum is mainly due to the Maxwell stress and settles around  $\alpha_M\simeq 10^{-2}$, a value that has also been obtained in 3D shearing box ambipolar simulations \citep[see Fig.~12 of][]{riols19}.  Figure~\ref{fig_alpha} shows that the laminar component of the Maxwell stress $\alpha_{M_L}$ contributes to a large fraction of the total stress  (see Eq.~\ref{eq_laminar_stress}), like in the box simulations. This term is actually associated with the laminar magnetic field that supports the wind structure.  {We find, however, that the turbulent stress is not negligible in the simulation, raising the question of the origin of such turbulence. Actually, despite the strong ambipolar diffusion, it can be seen that the disc is still  linearly unstable to magnetic perturbations (a form of damped MRI). The criterion for instability, given by \citet{kunz04}, for radially independent fluctuations and an unstratified disc is:
\begin{equation}
\dfrac{3\Omega^2}{k_z^2V_{A_z}^2}-1 > \dfrac{1}{\text{Am}^2}
,\end{equation}
where $k_z$ is the vertical wavenumber and $V_A$ is the vertical Alfven speed. The marginal case corresponds to $k_z=k_{z_{min}}\simeq 1/H,$ which gives the condition:
\begin{equation}
\label{eq_instab_Am}
\text{Am}>\text{Am}_c\simeq\left(\dfrac{3\beta}{2}-1\right)^{-1/2}
.\end{equation}
In our simulations, $\beta$ ranges from ten in the gaps to $10^4$-$10^6$ in the ring regions. The condition of instability is marginally satisfied in the gap regions and largely verified in the ring regions with $\text{Am}_c$ ranging from 0.008 to 0.0008. However, this criterion assumes that the field is purely vertical. The condition (\ref{eq_instab_Am}) can be generalised to a disc with a horizontal component of the magnetic field. In that case, the critical $\text{Am}_c$ has to be multiplied by a factor $V_A/V_{A_z}$ where $V_A$ is the total Alfven speed. In simulations, we typically find that the critical Am, taking into account this factor, is slightly less than one in the ring regions, allowing large-scale unstable modes to develop.}
\\

\subsection{Ring structures and origin}
\label{rings_origin}
As already mentioned in Section \ref{general_properties}, the disc develops radial density variations, meaning rings and gaps,  associated with zonal flows. These self-organised structures are not new in these types of simulations and were actually found in a variety of non-ideal MHD flows threaded by a net vertical field:  Hall-dominated turbulence \citep{kunz13}, ambipolar-dominated flows \citep[with or without ohmic diffusion, see][]{simon14,bai15}, and flows combining all non-ideal effects \citep{bai15,bethune17}. 
To look into these structures in more detail, in Fig.~\ref{fig_rhobz} (top) we show the time-averaged surface density as a function of radius. The two first ring-gap structures  in the inner disc are the most prominent ones and reach amplitudes $\Delta \Sigma \simeq 150\%$ of the mean background density $\Sigma_0$. The  following ring-gap structures  are fainter and their amplitudes are about $50\%$ to $60\%$ of the background density. Their typical {radial} size is about five times the disc scale height. Within the simulation time ($\sim 1000$ orbits), these structures remain almost steady. Figure~\ref{fig_zonal_flow}  shows the configuration of the poloidal magnetic field lines in the inner disc at $t=350 T_0,$ while  Fig.~\ref{fig_rhobz} (bottom) shows the time and vertically averaged $B_z$ as a function of radius. We can clearly see that the vertical magnetic field is concentrated into very thin shells at the location of the gaps (minima of $\Sigma$). Outside these shells, the net vertical flux is close to zero. This  behaviour is reminiscent of other MHD simulations \citep[ideal or not, see][]{bai14b,bai15,suriano18,riols18} and the origin of the magnetic shell seems clearly connected to the formation of the ring. 

In Fig.~\ref{fig_streamlines}, we also show the instantaneous poloidal streamlines around the ring structures at $t=350 T_0$. For $z \lesssim 3.5 H$ (white dashed lines), the flow is mainly radial (accreting in the southern hemisphere and decreting in the northern hemisphere) and seems to follow the rings, moving in a wave-like manner. In particular, it plunges toward the midplane at the gap location. This is line with the poloidal velcocity measured by \citet{teague19} in $^{12}$CO lines in the disc surface around HD 163296. Above this altitude, the flow is mainly vertical (pointing outward) and dominated by the wind component. \\

One important question is to understand the origin of these structures (rings + magnetic shell)  when ambipolar diffusion  is the dominant non-ideal MHD effect. Recently, \citet{riols19} proposed that the rings appear spontaneously in wind-emitting discs and are the result of a linear and secular instability, driven by the MHD wind. The process works as follows: a small initial radial perturbation of density generates a radial flow directed towards the gaps, because of the radial viscous stress.  The magnetic field, initially uniform, is radially transported towards the gaps (by the radial flow) and the excess of poloidal flux induces a more efficient mass ejection in the gaps, which reinforces the initial density perturbation. In this instability model, the gas is then depleted vertically by wind plumes rather than being radially accumulated into rings. This process, however, has only been shown in a local shearing box model, and its persistence in global configurations remains to be demonstrated. In particular, global models include a net vertical stress and a net radial accretion velocity that were not present in the local {framework}. \\

We then explore how these rings form and evolve in the global configuration.  Firstly, in the snapshots of Fig.~\ref{fig_zonal_flow} and \ref{fig_streamlines}, it seems obvious,  in particular in the southern hemisphere, that the wind is not uniform and homogeneous in the radius. Actually, it is composed of wind plumes that emanate from the regions where the poloidal field is maximum (gap regions). We checked that streamlines connected to the gap regions have the largest vertical (outward) velocity component. 
In addition to that, in Fig.~\ref{fig_massbudget} (top) we show the {radial} divergence of the flow $1/r \,\,{\partial}/ \partial r  (r \overline{v}_r)$, averaged in time and in the vertical direction {in the disc},   as a function of the radius. We see that the divergence is positive within the rings and negative within the gaps. This implies that on average there is a {converging} radial flow oriented toward the gaps {in the disc}. In the bottom panel of Fig.~\ref{fig_massbudget}, we also show that the mass loss rate  $\dot{\sigma}_w$ computed at $z=\pm 5H$ (in the wind region)  is maximum at the gap location, at least for the first three gaps, so the wind removes material from the regions where the vertical magnetic flux is concentrated. All these elements indicate that the mechanism  studied by \citet{riols19} in the local framework also works in the global configuration. 

One major difference in the global configuration is that there is a net accretion flow $\int_{-z_d}^{z_d} v_R dz /(2z_d)$ (with $z_d=3.5H$), which varies  from 0.002 $v_{K_0}$ at  $R=15 AU$ to 0.01 $v_{K_0}$  at  $R=100 AU$. This accretion flow is located mainly at the disc's southern surface (see Figs.~\ref{fig_zprofile} and \ref{fig_streamlines}) and should transport the magnetic flux radially over several AU. However, we do not see any motion of the magnetic structures during the course of the simulation; they rather stay at a fixed position. To better understand this, we examine the induction equation for $-\overline{B}_\theta \simeq \overline{B}_z$
\begin{equation}
\label{eq_induction_Bz}
\dfrac{\partial (-\overline{B}_\theta)}{\partial t} =   \dfrac{1}{r} \dfrac{\partial}{\partial r} \left[r\overline{E}_\phi  \right]=\dfrac{1}{r} \dfrac{\partial}{\partial r} \left[r (\overline{u_r B_\theta} - \overline{u_{\theta} B_r} + \overline{\eta_A J_\perp})\right ]
.\end{equation}
Figure~\ref{fig_induction} shows the radial profile of each term in the right-hand side of the induction equation. We see that the term related to the vertical stretching of the field $\frac{1}{r} \frac{\partial}{\partial r} (-r\overline{u_{\theta} B_r})$ plays little role in the evolution of $\overline{B}_z$. The latter is rather controlled by the two other terms, meaning radial transport by $u_r$ and ambipolar diffusion. Clearly, Fig.~\ref{fig_induction} shows that these two terms cancel each other out, the first one tends to transport the flux inward ($\frac{1}{r} \frac{\partial}{\partial r} (r\overline{u_r B_\theta})>0$ to the left of the $B_z$ maxima), while the ambipolar term seems to diffuse the field outward ($\frac{1}{r} \frac{\partial}{\partial r} (r\overline{\eta_A J_\perp})>0$ to the right of the $B_z$ maxima). Therefore, ambipolar diffusion plays a key role in the magnetic flux transport and seems to prevent the magnetic shells associated with the ring structures from being accreted inwards. We note that the mass is mainly accreted at the disc surface, so the density-weighted radial {velocity remains small. Hence, rings are expected to be advected at the mass accretion speed, which is much smaller than the gas velocity at the disc surface. Because our models are integrated only for $1000\,T_0$, this effect cannot be directly checked in the simulations.}

{Finally, we remind  the reader that our simulations are axisymmetric and are thus unable to capture non-axisymmetric effects such as the Rossby wave instability (RWI), which might be important in the evolution of the pressure maxima. We of course checked that minima of potential vorticity $(\nabla \times \mathbf{v})_{z}/\Sigma$ are located in the ring regions and could in principle excite the RWI \citep{lovelace99}. Nevertheless, the axisymmetry and stability of the radial structures seems to be preserved in 3D non-axisymmetric simulations  \citep{bethune17,suriano18b}. One possibility is that the RWI is stabilised by the turbulent stress or by the strong toroidal field \citep{yu09,gholipour15}. Whether these stabilising effects are important or not in 3D simulations remains, however, to be demonstrated. } 

\section{Dust dynamics}
\label{sec_dust}
\begin{figure*}
\centering
\includegraphics[width=\textwidth]{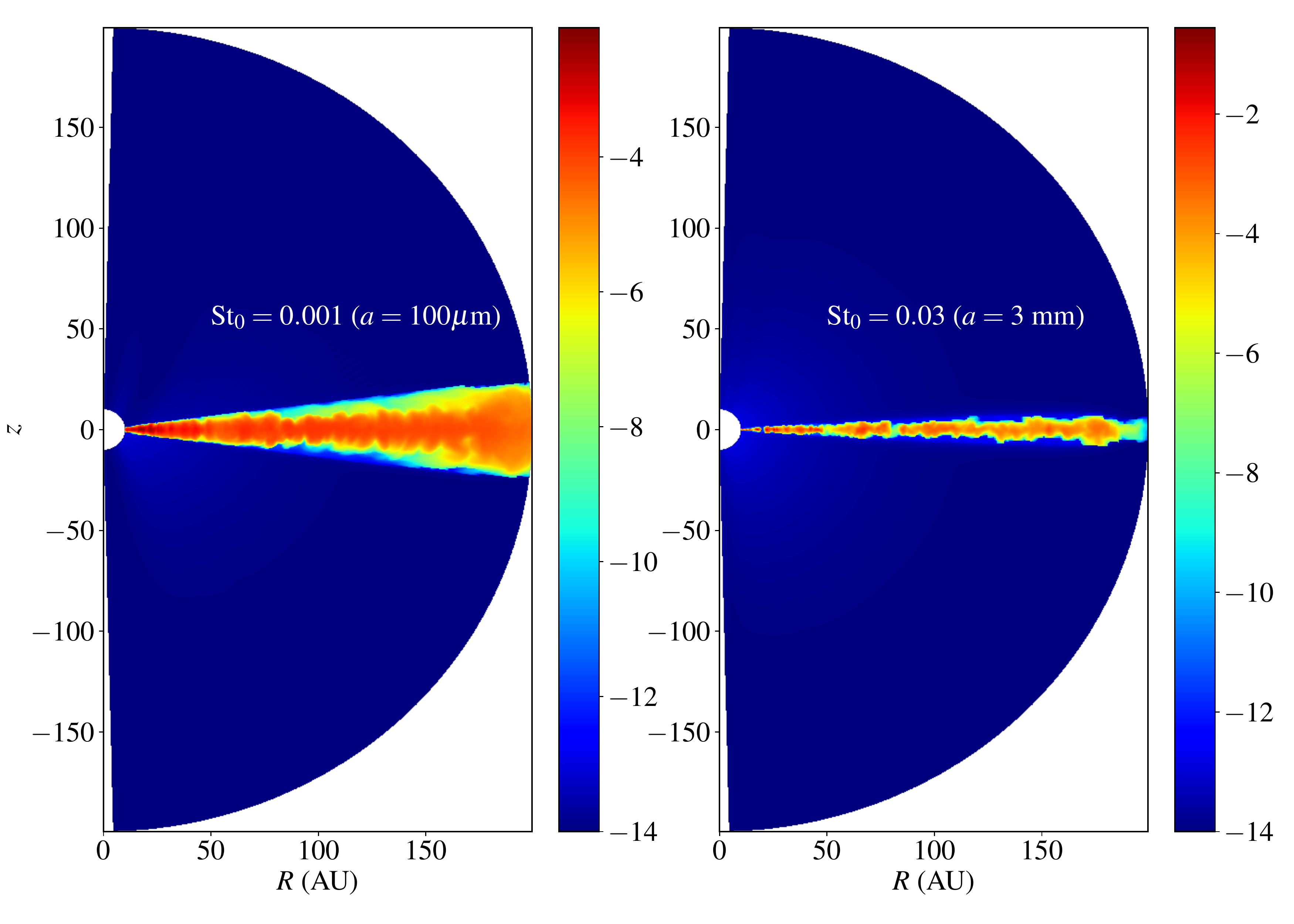}
 \caption{Dust density distribution for grains of size  $100 \,\mu$m (left) and $3$ mm (right), for $\beta=10^3$. This is averaged in time between 400 and 650 inner orbits}
\label{fig_rhodust}
\end{figure*} 
\begin{figure*}
\centering
\includegraphics[width=\textwidth]{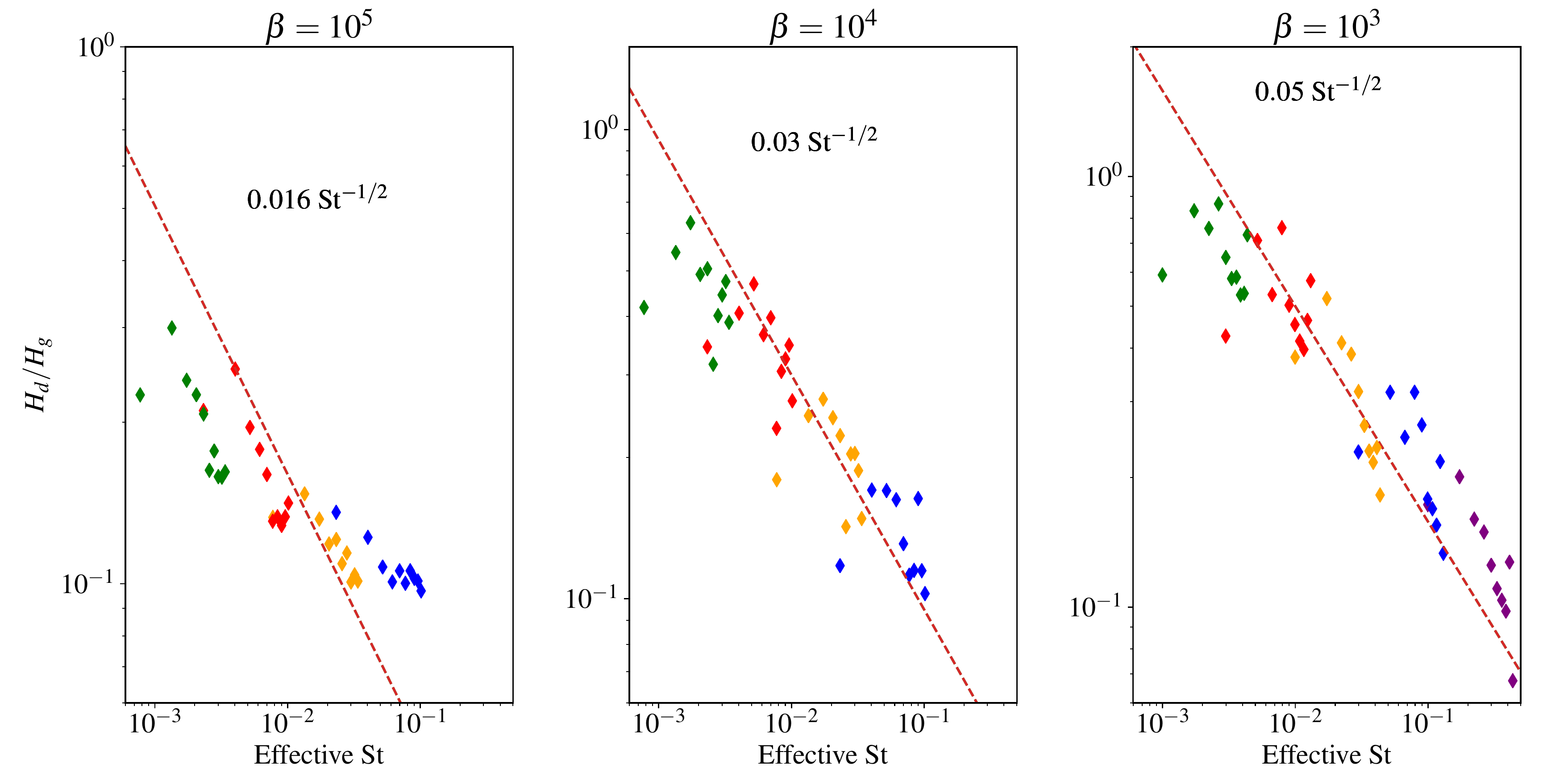}
 \caption{Dust to gas scale height ratio as a function of the effective Stokes number for $\beta= 10^5$ (left), $\beta= 10^4$ (centre) and $\beta= 10^3$ (right). Each colour accounts for a given particle size (purple: 1cm, blue: 3 mm, yellow: 1mm, red: 0.3 mm and green: 0.1mm.)}
\label{fig_Hd}
\end{figure*}

\subsection{Initialisation}

We now explore the dust dynamics in the non-ideal windy MHD flows presented in Section \ref{sec_without_dust}. We performed simulations for different $\beta$ and different grain sizes ranging from $a=100\, \mu m$ up to $a=1 $ cm. This corresponds to Stokes number St$_0$ (at $z=0$ and $R=10$ AU)  from 0.001 to 0.1 (see Section \ref{conversion} for the conversion between sizes and Stokes numbers). In total, we ran four different simulations: the two first at $ \beta=10^5$ and $ \beta=10^4$ both contain $0.1, 0.3, 1,$ and $3$ mm grains (respectively, St$_0$ of 0.001, 0.003, 0.01 and 0.03). The third simulation is for  $\beta=10^3$ and contains $0.1, 0.3, 3,$ and 10 mm grains (respectively, St$_0$ of 0.001, 0.003, 0.03 and 0.1). The last simulation at $\beta=10^3$ contains a unique species of 1 mm in size (St$_0$=0.01). 

The initial conditions for the dust are similar for all $\beta$ and Stokes numbers. The dust velocity is initially Keplerian with zero perturbation. The density at $t=0$ is
\begin{equation}
\rho_d(t=0)= \rho_{d_0} \left(\dfrac{R}{R_0}\right)^{-3/2}\, \exp\left[\dfrac{4GM}{c_s^2} \left( \dfrac{1}{\sqrt{R^2+z^2}}-\dfrac{1}{R}\right)\right]
,\end{equation}
with the initial $H_d/H_g= 0.5$. $\rho_{d_0}$  being fixed so that the ratio of surface densities $\Sigma_d/\Sigma$ is $0.0025$. In most of the simulations, we simultaneously integrate  four different species so that the total dust-to-gas density ratio is 0.01.  We note that for simplicity we use the same initial mass distribution for each species,  which is not likely to be the case in real protoplanetary discs. However, since the gas-to-dust ratio remains small when summed over all sizes,  this approximation has little effect on the results, {as we can later re-normalise the dust size distribution to whatever is needed. }
\begin{figure*}
\centering
\includegraphics[width=\textwidth]{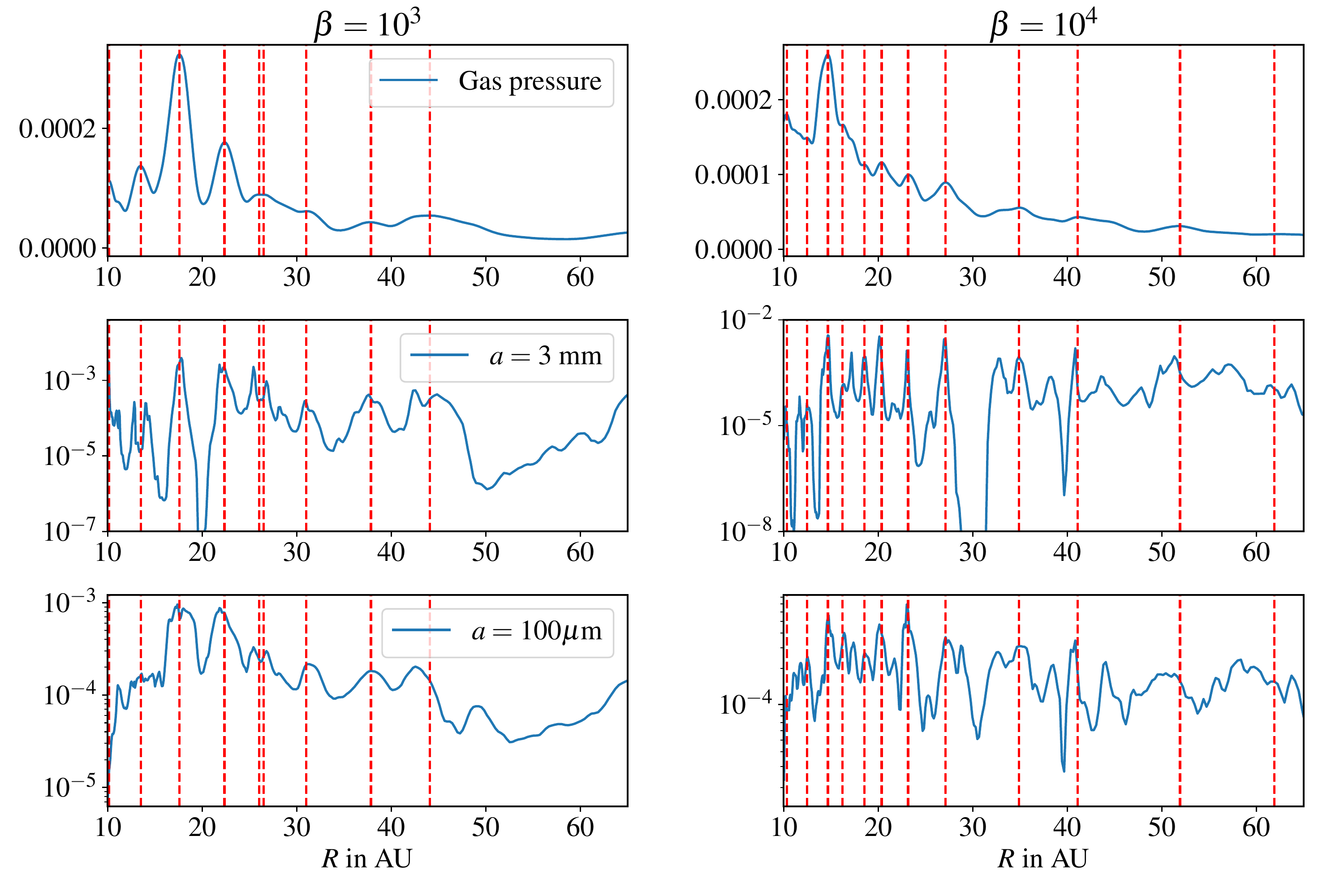}
 \caption{Top panels shows the gas pressure as a function of radius $R$ for $\beta=10^3$ (left) and $\beta=10^4$ (right). The central and bottom  panels show the dust surface density for  grains of  $3$ mm and $100 \,\mu$m, respectively. Quantities are averaged vertically between $\theta=\theta_-$ and $\theta=\theta_+$ ($z \pm 3.5 H)$ and in time  between 400 and 650 inner orbits.}
\label{fig_concentration}
\end{figure*} 
\subsection{Settling and dust scale height}

At the beginning of each simulation, large dust particles fall towards the midplane within a time proportional to $ \Omega^{-1}/\text{St}$ \citep{dullemond04}. This is the settling phase. 
As soon as turbulent diffusion and mixing become significant,  dust particles stop settling towards the midplane, and their mean vertical distribution reach an equilibrium. As we showed in Section \ref{conversion}, the midplane Stokes number St  depends on $R^{1/2}$, so the settling time is proportional to $R$ within the disc midplane.  For a 1mm particle (St$_0=0.01$), the typical settling time is about $100\, \Omega_0^{-1}$ at R=10 AU and $2000\, \Omega_0^{-1}$ at R=200 AU (the disc's outer radius). This remains smaller than the computation time ($\simeq 4000 \,\Omega_0^{-1}$ or 650 inner orbits).  However, for smaller particles, typically with St$_0=0.001$, the settling time is longer than the simulation time at large radii,  so the vertical dust distribution at these locations might not be in equilibrium and may still be evolving slowly in time. 

Figure~\ref{fig_rhodust} shows the typical dust density distribution in the disc averaged between 400 and 650 inner orbits, for $\beta=10^3$ and two different particle sizes corresponding to $a=100 \mu$m (St$_0=0.001$) and $a=3$mm (St$_0=0.03$). Clearly, the dust layer is thinner for the highest Stokes number, which is physically expected, since large particles are more sensitive to gravitational settling. From these density maps, it is possible to calculate the typical dust scale height as a function of the radius $R$. 
To be quantitative, we define the dust scale height $H_d\,(\text{St}_0,\beta,R)$ as the altitude $z,$ in such a way that
\begin{equation}
\label{eq_defHd}
\langle \rho_d \rangle (z=H_d) = \langle \rho_d \rangle (z=0)   \, e^{-\frac{1}{2}} \simeq 0.6\,\langle  \rho_{d0} \rangle .
\end{equation}
We find that the dust to gas scale height ratio $H_d/H_g$ is a decreasing function of radius $R$. This is due to the fact that the effective Stokes number increases with $R^{1/2}$ (see Eq.~\ref{eq_conversion}). We note that the power law depends on the choice of the disc's surface density profile and its dependence on $R$.  For instance, for $\beta=10^4$ and a 1mm particle, we find that the dust layer size is $0.26 H_g$ (0.13 AU) at $R=10 AU,$ and $0.18 H_g$ (0.9 AU) at $R=100 AU$.

For each $\beta$ and particle size, we divide the radial domain into ten bins (or parcels) and calculate the average ratio $H_d/H_g$ and average Stokes number within each parcel of the disc. In Fig.~\ref{fig_Hd}, we show the resulting $H_d/H_g$ as a function of the average `effective' Stokes number, where each colour accounts for a different particle size (or fixed St$_0$). 
As expected, we find that the ratio $H_d/H_g$ decreases with the Stokes number, and we end up with the following scaling relations:
\begin{align}
\label{eq_Hd1}
H_d/H \simeq 0.016 \,\, \text{St}^{-\frac{1}{2}}   \quad  \text{for} \quad \beta = 10^{5},\\ \simeq 0.03 \,\, \text{St}^{-\frac{1}{2}}   \quad  \text{for} \quad \beta = 10^{4},\\ \simeq 0.05 \,\, \text{St}^{-\frac{1}{2}}     \quad  \text{for} \quad \beta = 10^3.
\end{align}
We note that for the largest particle size (St$_0=0.1$ for $\beta=10^3$ or St$_0=0.03$ for $\beta=10^5$), we lack resolution since the dust is contained within two grid cells around the midplane. In that case, the ratio $H_d/H_g$ is probably overestimated. This could explain why the blue points in the left panel of Fig.~\ref{fig_Hd} lie above the expected scaling law (dashed line).  

The dependence on St$^{-1/2}$ is actually reminiscent of many other disc simulations coupling dust and gas turbulence \citep{fromang06b,zhu15,riols18}.  It can be obtained using the classical diffusion theory \citep{morfill85,dubrulle95} in which the equilibrium distribution in the vertical direction results from the balance between the gravitational settling and the diffusion due to the turbulence. In that theory, the turbulence is assumed to be homogeneous and isotropic, and can be described through a uniform, effective diffusion coefficient. 
We note that the local simulations of \citet{riols18} found a deviation from the simple power law St$^{-1/2}$ at $\beta=10^3$ and for small particles. In our case, we do not see such a deviation. Moreover, for $\beta=10^4$ and $\beta=10^3$, the ratio $H_d/H_g$ is two times larger than those measured in the local shearing box (see comparison with Fig.~7 of \citet{riols18}). This is due to a larger vertical rms turbulent velocity measured in the midplane in global simulations ($v_{z_{\text{rms}}} \sim 0.07 c_s$) compared to local simulations (where $v_{z_{\text{rms}}} \sim 0.04 c_s$, see Fig.~2 of \citet{riols18} for $\beta=10^3$).
\begin{figure*}
\centering
\includegraphics[width=\textwidth]{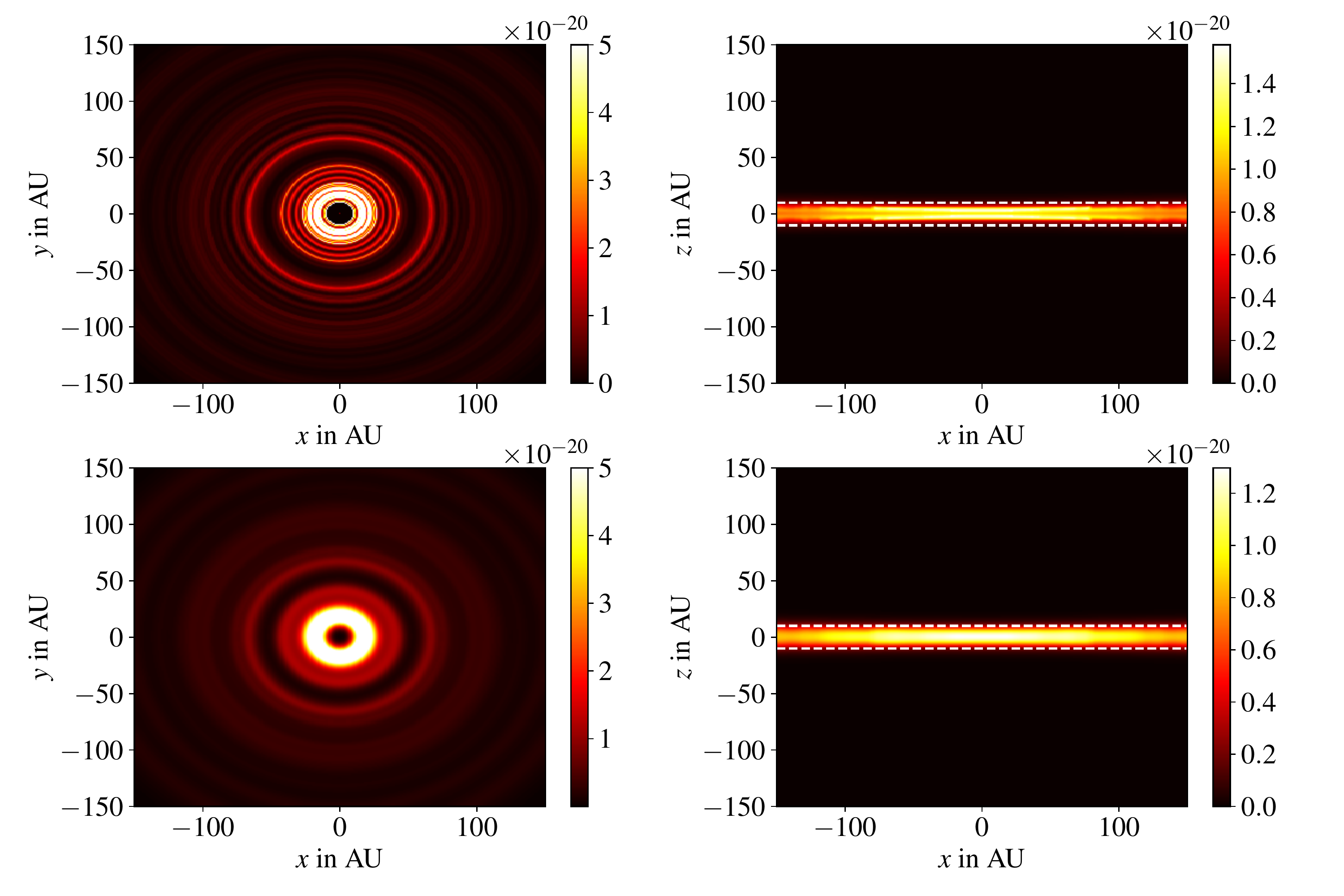}
 \caption{Top: Flux density $\nu F_\nu$ in W/m$^2$/pixel calculated by MCFOST, with the disc viewed face-on (left panel) and edge-on (right panel) for $\lambda=1$ mm. This is calculated for $t=500 T_0$ and $\beta=10^3$. Bottom: same images but convolved with a Gaussian kernel with a full width half maximum (FWHM) of 3 AU. The white dashed lines in the right panels shows the surface $z=H_g$ for $R=200$ AU.}
\label{fig_mcfost1}
\end{figure*} 
\begin{figure}
\centering
\includegraphics[width=\columnwidth]{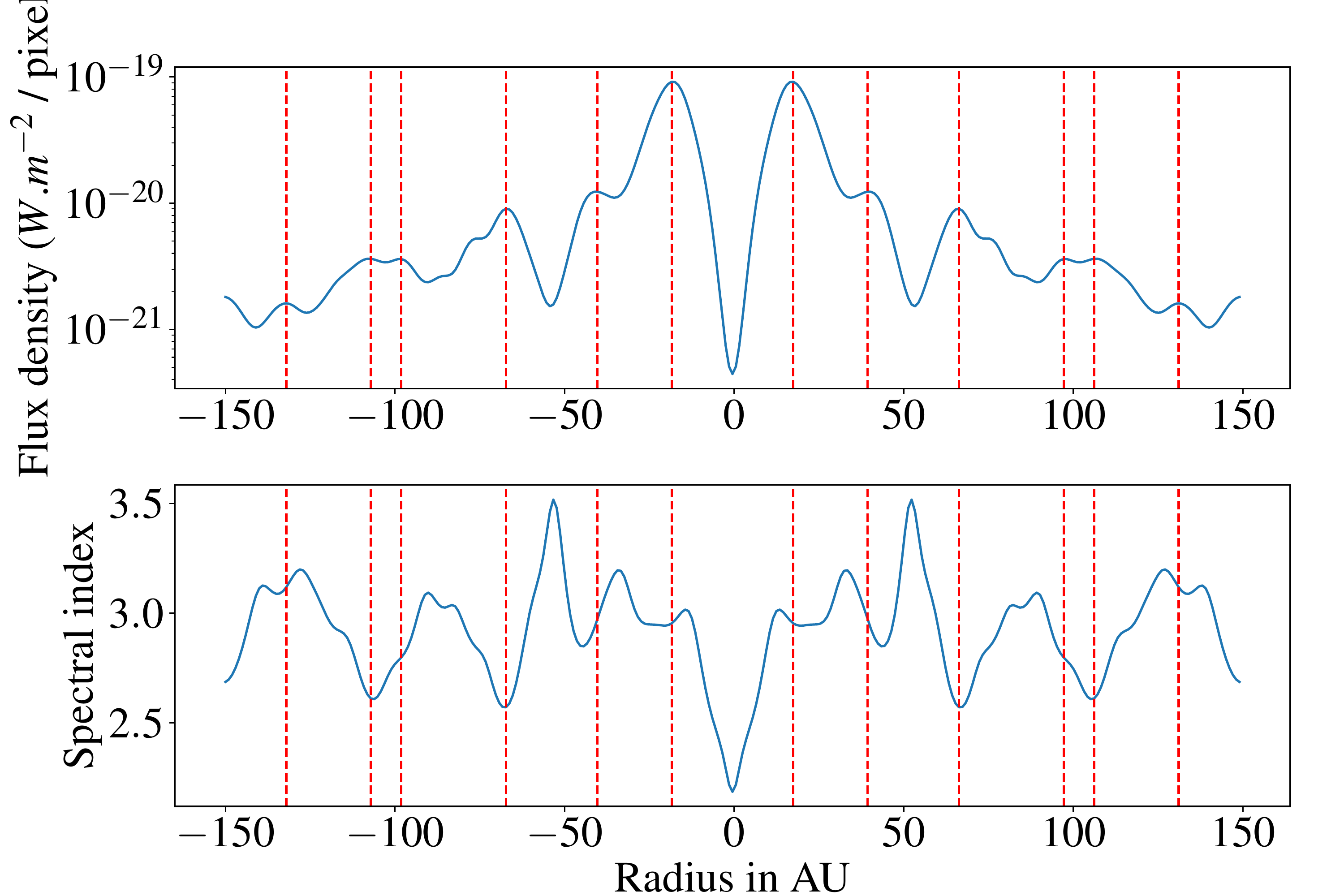}
 \caption{Top:  Convolved flux density $\nu F_\nu$ in W/m$^2$/pixel as a function of radius. Bottom: Spectral index $\alpha_s = d \log F(\nu)/d \log \nu$ as a function of radius measured between $\lambda =1$ mm (ALMA band 7) and $\lambda =3$ mm (ALMA band 3) }
\label{fig_mcfost2}
\end{figure} 
\begin{figure*}
\centering
\includegraphics[width=\textwidth]{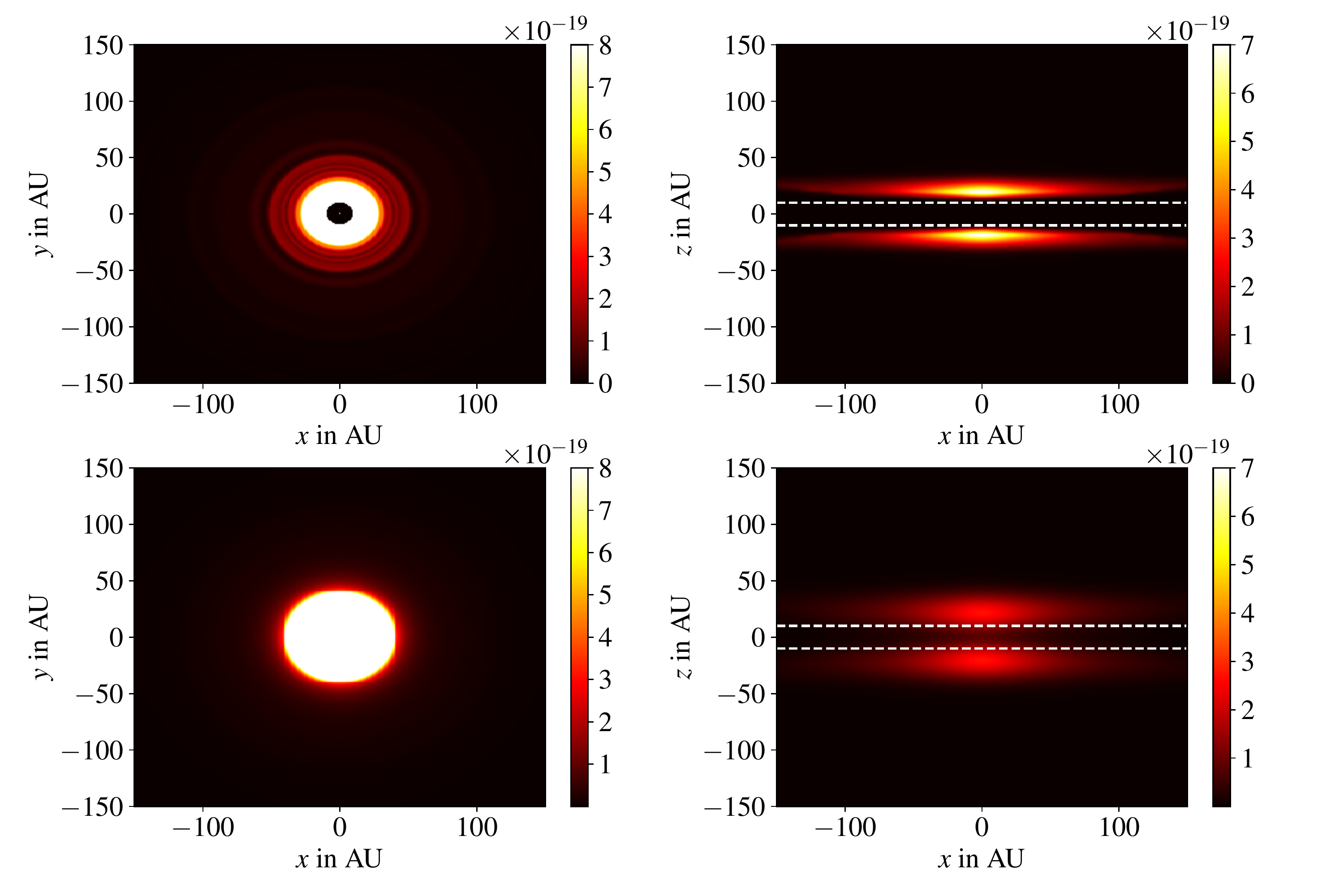}
 \caption{Top: Flux density $\nu F_\nu$ in W/m$^2$/pixel calculated by MCFOST, with the disc viewed face-on (left panel) and edge-on (right panel) for $\lambda=1.65$ microns. This is calculated for $t=500 T_0$ and $\beta=10^3$. Bottom: Same images but convolved with a Gaussian kernel with a full width half maximum (FWHM) of 10 AU. The white dashed lines in the right panels shows the surface $z=H_g$ for $R=200$ AU.}
\label{fig_mcfost3}
\end{figure*} 

\subsection{Concentration in pressure maxima}

As we showed in Section \ref{rings_origin}, rings of matter arise spontaneously in wind emitting discs, and they correspond to pressure maxima in the locally isothermal case. These maxima are then the locations for dust accumulation and could play an important role in the formation of planetesimals.  The concentration of dust is, however, only possible if the drift velocity, proportional to St and the local pressure gradient, is larger than the velocity associated with the radial turbulent diffusion. We then expect concentration of dust into radial bands  to be more efficient for sufficiently large values of the Stokes number. In Fig.~\ref{fig_concentration}, we show the radial profiles of the gas pressure and dust surface density $\Sigma_d$, averaged in time, for two different plasma beta parameters and two different grain sizes. For $\beta=10^3$, both large and small grains concentrate into pressure maxima (with position delimited by the red dashed vertical lines). However, 3mm grains, the contrast in density, at some radii, can be 200 to 500 times larger than the contrast measured for 0.1 mm grains.  Although the density contrast is pretty high for $a=3$ mm, the dust-to-gas ratio remains bounded and smaller than 0.04. We also note that the radial dust structures are thinner in the case of larger particles. For $\beta=10^4$, the results are similar, except that the pressure maxima are less pronounced and the dust is less concentrated on average. This is particularly obvious for 0.1 mm grains, for which the contrast in density never exceeds ten. Also, the typical separation between the dust rings in the case $\beta=10^4$ is about 2.5 $H$ at $R\simeq20$ AU, while it is about 4 $H$ for $\beta=10^3$. {By fitting the radial density profiles with gaussian functions, we find that the half-width of the dust rings at $R=20 AU$ is $\sim 0.3-0.4$ AU (0.3-0.4 $H$) for $\beta=10^3$ and $a=3$ mm, while it is $\sim 0.15$ AU for $\beta=10^4$. For comparison, the pressure bumps in the gas have a typical half-width of 1.2 $H$.}

 We note that for $\beta=10^5$, a few rings form, but they are much fainter than in the case  $\beta=10^4$ or $\beta=10^3,$ and therefore we do not see any clear radial structures in the dust. We cannot rule out  that more prominent structures could form if the simulation were run for a longer time period. 

\section{Radiative transfer with MCFOST}
\label{sec_mcfost}

The aim of this Section is to produce synthetic images derived from our simulations in order to predict the observable properties of ambipolar-dominated discs threaded by a net vertical field. For that we use the 3D continuum and line radiative transfer code MCFOST based on the Monte Carlo method \citep{pinte06}.  

Assuming a given gas and dust density distribution, the code first computes the temperature structure assuming that the dust is in radiative equilibrium with the local radiation field. This is done by propagating photon packets, originally emitted by the central star (assumed to be of solar-type), through a combination of scattering, absorption and re-emission events until they exit the computation grid. During this step, we use 1 280 000 photon packets in total.  It is assumed that the dust opacities are fixed and do not depend on the temperature during this stage. 
Observables' quantities (images) are then obtained via a ray-tracing method, which calculates the intensities by integrating the radiative transfer equation along rays perpendicular to each pixel of the image. 

The disc extends from an inner cylindrical radius $R_{in}=10$ AU to an outer limit $R_{out}=200$ AU, with a total gaseous mass equal to $5\times 10^{-2}$ solar mass.  Dust grains are defined as homogeneous and spherical particles (Mie theory) with sizes distributed according to the power law $dn(a)/da=a^p$ with $p=-3.5$. 
During the radiative transfer calculation, we consider 100 different sizes of grains ranging from 0.03 $\mu m$ to 1 cm.  We note that our PLUTO simulations only contain four dust species of $0.1, 0.3, 3$ and 10 mm, which all possess the same initial mass. Thus, 
to construct the distribution of each grain population, MCFOST automatically interpolates the axisymmetric PLUTO density fields between the input sizes, and then normalises them to make sure that $dn(a)/da=a^p$ with the additional constraint that the gas-to-dust ratio is 100. Such a re-normalisation  procedure can be achieved because the dust has almost no feedback on the gas (since the dust-to-gas ratio remains low in all simulations).   We note that during the interpolation, we assume that the smallest grains of 0.03 $\mu m$ follow the gas distribution exactly. \\

In the upper panels of Fig.~\ref{fig_mcfost1}, we show the image of our disc for $\beta=10^3$ calculated at a given wavelength  ($\lambda = 1$ mm), corresponding to ALMA band 7, in a face-on (left) and edge-on (right) configuration. To take into account the resolution of instruments like ALMA, in the lower panels we show the same images convolved with a Gaussian kernel with a full width half maximum (FWHM) of 3 AU. This resolution is typically expected for an object located at 140 pc and for a radio-telescope baseline of 15km. In the original image, between ten and 50 AU we see a series of eight concentric rings that correspond to the location of the pressure maxima in Fig.~\ref{fig_concentration} where the millimetre dust is trapped. In the convolved image, these rings are reduced to one bright inner ring and a second fainter ring (see also top panel of Fig.~\ref{fig_mcfost2}). A wide gap is obtained at 50 AU, followed by other bright rings in the outer regions (in particular one at $R=70$ AU and another at $R=100$ AU in the convolved image).  The typical contrast at 50 AU in the convolved image is 7-8, while the separation between the gaps or the rings is close to 20 AU. In the edge-on configuration, the disc thickness is about 5 AU at $R=200$ AU, which corresponds to 0.5 $H_g$. For $\lambda = 1$ mm, most of the thermal emission comes from grains of $\lambda/2\pi\simeq 160 \mu$m, which are indeed contained in a layer of  $\approx0.5 H_g$ at $R=200$ AU according to Fig.~\ref{fig_Hd}. We note that the emission is lower in the midplane of the disc than in the layers above (see Appendix \ref{appendixA} showing the vertical profile of the flux density). This is due to the smaller temperature in the midplane, resulting from the strong sedimentation of mm to cm grains. 

We also show, in the bottom panel of Fig.~\ref{fig_mcfost2}, the spectral index $\alpha_s = d \log F(\nu)/d \log \nu$ as a function of the radius obtained by comparing the image at $\lambda = 1$ mm with the image at $\lambda = 3$ mm. The spectral index ranges between 2.5 and 3.5, suggesting that the disc deviates slightly from a black body (whose spectral index is 2 in the Rayleigh-Jeans regime). The spectral index is found to be at a maximum in the gap regions and at a minimum within the rings, in agreement with the fact that the surface density, and therefore the optical thickness, is lower in the gaps.  

Finally, we computed the scattered light emission from micron-size dust at the mid-infrared wavelength $\lambda=1.65$ microns, which corresponds to band H of the SPHERE instrument. The synthetic images of the discs viewed face-on and edge-on are shown in  the top panels of Fig.~\ref{fig_mcfost3}. In the face-on view, we distinguish several ring structures, although they remain much fainter than in the ALMA bands.  In the edge on view, the emission comes mainly from the surface $z \simeq 2H$. Since the disc is optically thick, the infrared emission from the midplane region is totally absorbed, allowing only the detection of scattered light from the outer edges of the upper and lower surfaces of the disc. To take into account the beam of the instrument, we convolve the original images (top panels) with a Gaussian kernel of FWHM of 10 AU (bottom panels). We find that rings are absent in the convolved images, since their actual size is smaller than the beam size.

\section{Conclusion and discussion}
\label{sec_conclusions}

In summary, we investigated the gas and dust dynamics of protoplanetary discs threaded by a large-scale magnetic field and subject to ambipolar diffusion. Using radially and vertically stratified global axisymmetric MHD simulations, we showed that accretion occurs mainly at the disc surface and is mainly driven by the laminar torque associated with large-scale MHD winds. For plasma $\beta$ of $10^3$, the accretion rate is about $10^{-7} M_\odot/\mathrm{yr}$ at $R=10$ AU. {We note that the mass loss rate represents a significant fraction of the mass accretion rate, so that $\mathrm{d}\log\dot{M}_\mathrm{acc}/\mathrm{d}\log R\simeq 0.6$, typical of magneto-thermal winds. Hence, if we assume winds exist down to the disc's inner radius $R_\mathrm{in}\simeq 0.1\,AU$, one expects $\dot{M}_\mathrm{acc}(R_\mathrm{in})\lesssim 10^{-8}\,M_\odot/\mathrm{yr}$, which is fully compatible with observed accretion rates onto YSOs. } Within the midplane, the disc displays gaseous ring-gap structures associated with zonal flows,  with a segregation between vertical magnetic flux and matter. These structures are stable during the simulation and are likely to be triggered by the same mechanism described in \citet{riols19}, which involves a wind instability.

 We showed that centimetre to millimetre dust particles are highly sedimented in the midplane with a typical scaleheight $H_d/H_g \lesssim 0.25$ for $\beta=10^4$. For $R=100$ AU, this corresponds to a dust layer of size $\lesssim 1$ AU. This result fits the observational constraint of \citet{pinte16} in HL Tau (ALMA) particularly well and is in contrast with ideal MRI turbulence simulations  for which the dust-to-gas scale height ratio is much larger than that inferred from observations \citep{johansen06,fromang06}.

  Another result is that the dust particles are trapped within the pressure maxima and form thin axisymmetric structures with high contrast (for mm to cm particles) and typical separation of 2.5 to 5 $H_g$ at $R=20$ AU, {wider separation being obtained in stronger field situations}. {The width of the millimetre dust rings is about $0.3 -0.4H$ for $\beta=10^3$. This is comparable with the width of dust ring structures observed by the DSHARP survey \citep{dullemond18} in objects like AS 209 (although the comparison is done at different radii, and we implicitly assume that the ratio width over $H$ does not depend on radius). We estimate that the typical dust mass contained inside a single ring at $R\simeq20$ AU is about 5 Earth mass. } Using the radiative transfer code MCFOST, we were able to produce synthetic images of the disc in the ALMA band (1mm) for $\beta=10^3$. These images, when convolved with a Gaussian kernel that accounts for the instrumental resolution,  show concentric rings and gaps with typical contrast $\lesssim 10$ and separation of 20 AU. The spectral index within the gaps goes up to 3.5, while it remains below 3 within the rings, {similarly to observed spectral indices in HL-Tau} \citep{alma15}. At a mid-infrared wavelength (scattered light emission), we found that instruments like SPHERE would be unable to detect the rings, due to their insufficient resolution. 
  
{We note that our simulations are limited to the region between 10 and 200 AU, but we expect the instability giving rise to the ring-gap structures to still work in the inner and densest regions where ohmic diffusion predominates (see  \citet{riols19}. This could explain, for instance, the dust rings observed in the TW Hydra disc on scales smaller than 10 AU \citep{andrews16}. }  \\

It is relevant to compare our work with recent simulations of ambipolar-dominated flow.  In particular, \citet{suriano18} showed that the segregation of vertical magnetic flux and matter is due to the reconnection of highly pinched poloidal fields. This requires a current sheet in the midplane and a  poloidal field that bends within it.   In our case, we only see pinched poloidal fields at the very beginning of the simulations ($t<100 T_0$) when the rings are not yet formed. At later times, the field takes the configuration depicted in Fig.~\ref{fig_zonal_flow},  similar to what was obtained by \citet{bethune17}. This configuration shows a break of symmetry, inappropriate regarding reconnection in the midplane. The cause of the breaking of symmetry in our case is yet unknown, but appears to be a robust result independent of the initial condition. Another comparison can be made with the recent local box simulations of \citet{riols18} who computed the dust and gas motions around a fixed radius $R=30$ AU.  We found that the properties of the gaseous and dusty rings (separation and depth) are comparable in the local and global frameworks. However the settling of the dust is more pronounced in the shearing box in particular for $\beta=10^4$  and $\beta=10^3$.   \\

{We remind the reader that the the choice of cooling timescale adopted in our simulations prevents the development of the vertical shear instability (VSI). Recent radiation hydrodynamic simulations \citep{stoll14,stoll16, flock17} suggest that the VSI should actually be present in the outer regions of protoplanetary discs and induce substantial diffusion of dust particles. We argue, however, that the physics of the VSI strongly depends on the thermodynamics, and that the flux-limiting diffusion used by the previous authors might not be a good approximation away from the midplane where the VSI triggers the strongest motions. Moreover, a recent study by \citet{cui19} investigating the interplay between MHD winds in PP discs with the VSI shows that the hydrodynamic instability is weakened by MHD effects in the moderate  magnetisation regime ($\beta = 10^3$).}\\

This work is also {connected to} planet formation theory. The {formation of} gaseous rings may indeed be a way to overcome the `radial-drift' barrier  (growth of grains to pebbles)  and the `fragmentation' barrier ( growth from planetesimals to planets)    \citep{birnstiel16}. In particular, the trapping of grains in pressure maxima 
 can hinder their radial migration, accelerate their growth, and prevent their fragmentation \citep{gonzalez17}. Gaseous rings associated with zonal flows appear to be an alternative to other trapping mechanisms such as vortices, whose stability and origin remain uncertain \citep{lesur09}. An interesting avenue of research would  be to better characterise the grain growth inside these ring structures. Also, the strong sedimentation in the midplane, especially for $\beta=10^5$ or $\beta=10^4$ could also reinforce the so-called pebble accretion \citep{ormel10,lambrechts12}, and therefore speed up the growth of massive cores.  \\

{Our results suggest that self-organisation is unavoidable in weakly ionised discs threaded by a large-scale poloidal field, provided that the field is strong enough. The question of the strength of the field threading these discs is therefore of central importance. Unfortunately, there is no direct measurement of the field strength at these distances. This is a rather difficult task since the field strength corresponding to a $\beta=10^3$ disc is approximately 15 mG at 10 AU, which is a relatively weak field. On the theory side, the most recent core-collapse models \citep{masson16} suggest that discs form with a constant large-scale field of 100 mG, meaning stronger but still comparable to our $\beta=10^3$ disc model. As the disc evolves, the field strength is also expected to evolve, but no theory is yet able to predict whether the field should diffuse away, or on the opposite accumulates towards the centre (see e.g. \citealt{guilet13,bai17}). In any case, this suggests that $\beta\simeq 10^2-10^4$ is well within the large-scale field strength expected for protoplanetary discs, and therefore that wind-driven self-organisation is most probably present in these objects.}

{In the end, the main question is to unambiguously identify which mechanism is responsible for the structures we see, whether it is induced by embedded planets, wind-driven self-organisation, or other secular gas instabilities. As pointed out in the introduction, finding planets in the gaps is not sufficient, since all of the mechanisms will accelerate planet formation in the rings thanks to the presence of local pressure maxima that trap growing grains. The presence of a deviation from the Keplerian rotation profile is not really useful either, since this is expected as a result of the radial geostrophic equilibrium of the disc, which is always satisfied, with or without planets. More specific signatures are therefore desperately needed.  A first potential signature would be the eccentricity of the gap/ring structures, which is expected to be zero in the case of wind-driven self-organisation, but to exceed 0.1 if a planet of mass $\gtrsim 0.1$ Jupiter mass opens the gap \citep{zhang18}. Such eccentricity might be difficult to measure with current facilities but feasible with the next generation of instruments. 
In the particular case of wind-driven self-organisation, the kinematics of the gas is peculiar, since the gas tends to radially converge towards the gap, before being ejected. At the disc surface, which is typically tracked by CO observations, the gas is typically flowing radially, 'bouncing' onto the rings (Fig.~\ref{fig_streamlines}). This is the kind of kinematics that can be tested on the sky \citep[see e.g.][]{teague19}. 

Our global simulations also predict that the typical separation between the rings increases with the disc magnetisation, a result that has been demonstrated, as well in the shearing box \citep{riols19}. We also expected the accretion rate and mass loss rate in the wind to increase with the magnetisation. If the rings are formed via MHD winds, we would then expect a correlation between the ring separation and the accretion rate. This relation could be tested in the current and future observations.

Ideally, one would also want to look for the presence of magnetic field concentration in the gaps. We expect a contrast in field intensity reaching several tens, implying a field strength of a few hundreds mG at 10 AU in the gaps. This might not be achievable with current instruments, but such a detection would undoubtedly demonstrate the reality of this kind of mechanism. }

\section*{Acknowledgements}
 This project has received funding from the European Research Council (ERC) under the European Union’s Horizon 2020 research and innovation programme (Grant agreement No. 815559 (MHDiscs)). This work was  performed  using  the  Froggy  platform  of  the  CIMENT HPC  infrastructure (https://ciment.ujf-grenoble.fr).

\bibliographystyle{aa}
\bibliography{refs} 

\appendix

\section{Visibility in ALMA bands}
\label{appendixA}
In Fig.~\ref{fig_visibility}, we show the normalised vertical profiles of  the flux density $\nu F_\nu$ computed in different ALMA bands with MCFOST in the case of a disc viewed edge-on. These profiles are calculated at the centre of the image ($x=0$) and are normalised to their maximum (longer wavelengths have lower emissions). We see that the width of the profiles decreases with the wavelength, since a longer wavelength traces a larger, more sedimented particle size. In the midplane region at $z=0$, we observe a gap in the emission corresponding to a darkening of the disc. The depth of the gap increases at lower wavelength. The darkening is due to large sedimented grains (of cm size) absorbing the light and obscuring the disc midplane.  They also contribute to lower the temperature in the midplane, and thus reducing the emission from this region. 
\begin{figure}
\centering
\includegraphics[width=\columnwidth]{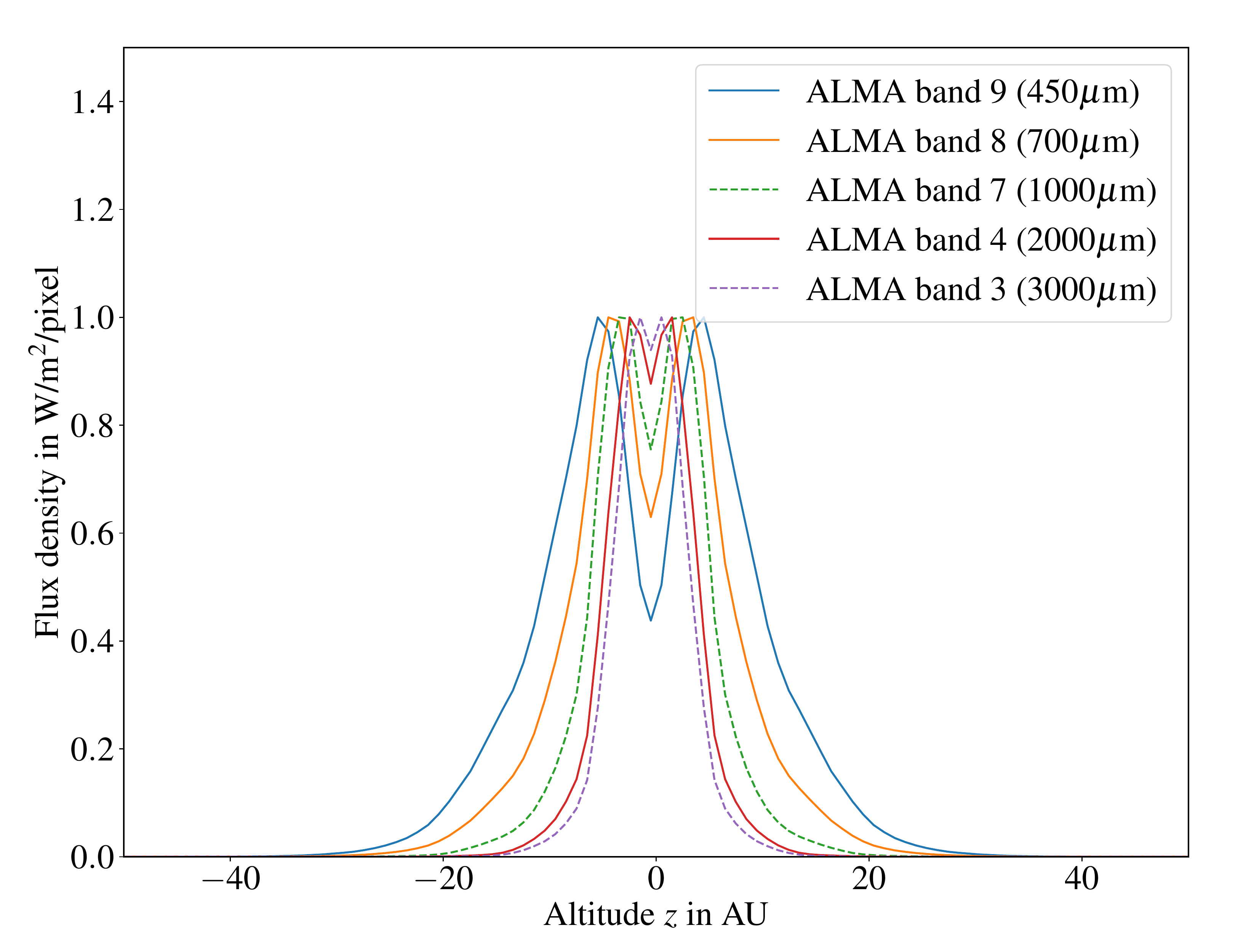}
 \caption{Vertical profiles of the flux density $\nu F_\nu$ for different wavelengths corresponding to different ALMA bands, computed with MCFOST in case of a disc viewed edge-on. The flux density is normalised to its maximum.}
\label{fig_visibility}
\end{figure} 

\end{document}